\documentclass[prd,superscriptaddress,amsfonts,amssymb,amsmath,showpacs,twocolumn]{revtex4-2}

\usepackage{bm}
\usepackage{amsfonts}
\usepackage{latexsym}
\usepackage[utf8]{inputenc}
\usepackage{graphicx}
\usepackage{amsmath}
\usepackage{palatino}
\usepackage{mathpazo}
\usepackage{textcomp}
\usepackage{multirow}
\usepackage{booktabs}
\usepackage{float}
\usepackage{booktabs}
\usepackage{dcolumn}
\usepackage{ragged2e}
\usepackage{hyperref}
\hypersetup{colorlinks,citecolor=blue}
\hypersetup{colorlinks=true,linkcolor=red,filecolor=magenta,    urlcolor=blue}
\usepackage{amsmath}
\usepackage{xcolor}
\usepackage{orcidlink}
\usepackage[caption=false]{subfig}
\usepackage{commath}
\def\jnl@style{\it}
\def\aaref@jnl#1{{\jnl@style#1}}

\def\aaref@jnl#1{{\jnl@style#1}}

\def\aj{\aaref@jnl{AJ}}                   
\def\apj{\aaref@jnl{ApJ}}                 
\def\apjl{\aaref@jnl{ApJ}}                
\def\apjs{\aaref@jnl{ApJS}}               
\def\apss{\aaref@jnl{Ap\&SS}}             
\def\aap{\aaref@jnl{A\&A}}                
\def\aapr{\aaref@jnl{A\&A~Rev.}}          
\def\aaps{\aaref@jnl{A\&AS}}              
\def\mnras{\aaref@jnl{Mon.~Not.~Roy.~Astron.~Soc.}}             
\def\prd{\aaref@jnl{Phys.~Rev.~D}}        
\def\plb{\aaref@jnl{Phys.~Lett.~B}}        
\def\prc{\aaref@jnl{Phys.~Rev.~C}}  
\def\prl{\aaref@jnl{Phys.~Rev.~Lett.}}    
\def\qjras{\aaref@jnl{QJRAS}}             
\def\skytel{\aaref@jnl{S\&T}}             
\def\ssr{\aaref@jnl{Space~Sci.~Rev.}}     
\def\zap{\aaref@jnl{ZAp}}                 
\def\nat{\aaref@jnl{Nature}}              
\def\aplett{\aaref@jnl{Astrophys.~Lett.}} 
\def\apspr{\aaref@jnl{Astrophys.~Space~Phys.~Res.}} 
\def\physrep{\aaref@jnl{Phys.~Rep.}}      
\def\physscr{\aaref@jnl{Phys.~Scr}}       
\def\commat{\aaref@jnl{Comm.~Math.~Phys.}}              
\def\science{\aaref@jnl{Science}}               
\def\cqg{\aaref@jnl{Classical Quant.~Grav.}}            
\def\jpcs{\aaref@jnl{JPCS}}                                     
\def\ijmpd{\aaref@jnl{Int.~J.~Mod.~Phys.~D}}                    
\def\grg{\aaref@jnl{Gen.~Relat.~Gravit.}}               
\def\rpp{\aaref@jnl{Rep.~Prog.~Phys.}}          
\def\npa{\aaref@jnl{Nucl.~Phys.~A}}        
\def\lrr{\aaref@jnl{Living Rev.~Rel.}}                   
\def\jcap{\aaref@jnl{J.~Cosmology Astropart.~Phys.}}    
\def\rmp{\aaref@jnl{Rev.~Mod.~Phys.}}   
\def\epjc{\aaref@jnl{Eur.~Phys.~J.~C}}

\allowdisplaybreaks[1]
\renewcommand{\arraystretch}{1.1}
\begin{document}
\color{black}       
\title{\bf Quark Stars in Ricci-Determinant Gravity with an Interacting Quark Equation
of State}
\author{Loreany F. Ara\'ujo  \orcidlink{0000-0003-1749-8354}}
\email{loreanyfa@alumni.usp.br}
\affiliation{ Departamento de Astronomia, Universidade de S\~ao Paulo\\
Rua do Mat\~ao, 1226 -- 05508--900, S\~ao Paulo, SP, Brazil\\
}

\author{Krishna Pada Das  \orcidlink{0000-0003-4653-6006}}
\email{krishnapada0019@gmail.com}
\affiliation{ Department of Mathematics, Indian Institute of Engineering\\
Science and Technology, Shibpur, Howrah-711 103, India\\
}


\date{\today}

\begin{abstract}
In the present study, we explore the fundamental properties of static, spherically symmetric quark stars composed of quark matter with an interacting quark equation of state (EoS) within the framework of Ricci-Determinant gravity. To this end, we adopt the relativistic stellar structure equations for compact objects derived in the literature. Our primary objective is to investigate deviations from General Relativity (GR) in key physical characteristics, particularly the mass--radius relation and stability criteria, arising from the free parameters of this extended gravitational theory. We see that, unlike the hadronic case, the model predicts a reduction in the compactness of quark stars. This parameter is also sensitive to gravitational binding-energy analysis, revealing a breakdown of the assumed universality. Furthermore, the formation of objects with high central densities is restricted by the instability conditions that arise when the contribution of perturbative terms exceeds by approximately half the contribution of ordinary GR, indicating a clear limitation in the theory.
\\
{\bf Keywords:} modified gravity, interacting quark matter, mass-radius relation, dynamical stability.
\end{abstract}

\maketitle

\section{Introduction}

In recent years, substantial attention has been directed toward extensions of GR, as the theory faces several conceptual and observational challenges. Enduring theoretical problems such as the emergence of spacetime singularities, the missing mass phenomenon commonly attributed to dark matter, the late-time cosmic acceleration inferred from observations of distant Type Ia supernovae (dark energy), and the lack of a consistent quantum formulation of gravity provide strong motivation for the development of modified or generalized gravitational theories. Moreover, an increasing body of astrophysical and cosmological observations further underscores the necessity of exploring gravitational frameworks that extend beyond GR \cite{Planck:2018vyg,Riess:1998cb,Ishak:2018his,Nojiri:2017ncd,Heisenberg:2018vsk,Clifton:2011jh}. GR is inherently a metric theory of gravity, and most proposed extensions are formulated within this geometric framework. Over the past few decades, considerable efforts have been devoted to modifying or generalizing GR in order to address various theoretical and observational challenges. In its standard formulation, gravity is understood as the manifestation of spacetime curvature, generated by the distribution of matter and energy through the energy–momentum tensor $T_{\mu\nu}$. Therefore, modifications to GR may be implemented either by extending the geometric sector (through curvature-based corrections), by altering the matter sector (via modifications of the energy–momentum contributions), or by introducing nonminimal couplings between matter and geometry.\\
Over the past decades, extensions based on the Palatini (or metric-affine) formulation have attracted considerable attention \cite{Olmo:2009xy}. Within the Palatini approach, the spacetime metric $g_{\mu\nu}$ and the affine connection $\Gamma^{\sigma}_{\mu\nu}$ are regarded as independent dynamical variables. Importantly, the connection is not assumed a priori to be the Levi-Civita connection. For standard GR, varying the action with respect to both the metric and the connection typically leads to conditions that constrain the connection to coincide with the Levi-Civita one. In this framework, the Einstein–Hilbert (E-H) action involves only first-order derivatives of the connection and directly reproduces the Einstein field equations, without the necessity of adding extra boundary terms. In contrast, the conventional metric formulation of GR includes second-order derivatives of the metric in the action, which makes the introduction of an appropriate boundary term essential for establishing a well-defined variational principle. In this regard, one may consider the well-known $f(R)$ gravity, in which the modification arises purely from the geometric sector by introducing scalar curvature invariants into the E-H action \cite{Nojiri:2006ri}. Furthermore, within this class of extensions, one can also study more general theories such as $f(g^{\mu\nu}R_{\mu\nu}, R_{\mu\nu}R^{\mu\nu})$ gravity \cite{Olmo:2011uz}, where the Lagrangian is constructed from powers and traces of the Ricci tensor, thereby providing a natural extension of GR. In this connection, readers may also explore Gauss–Bonnet gravity and its modification \cite{Zwiebach:1985uq,Nojiri:2005jg}, another extension of GR obtained by incorporating higher-order curvature invariant terms into the E-H action.\\

Now, if we examine Eddington gravity \cite{Banados:2010ix} and Eddington-inspired Born–Infeld gravity \cite{Jimenez:2017uyn}, it becomes clear that modified gravity need not be constructed solely through polynomial curvature invariants derived from contractions of the curvature tensor, especially the Ricci tensor. Instead, these theories formulate the gravitational action in terms of the determinant of the Ricci tensor. Motivated by these developments, Azri $et~al$. \cite{Azri:2021det},  proposed another extension of GR, known as Ricci-determinant (RD) gravity, constructed by incorporating the determinant of the Ricci tensor within the Palatini formalism. In this framework, the gravitational action is generalized by introducing an arbitrary function $f(\boldsymbol{D})$, where $\boldsymbol{D}$ denotes an invariant scalar quantity, and is expressed as $\boldsymbol{D}= |\textbf{det.}\mathcal{R}|/|\textbf{det.}g|$. The importance of this gravity theory stems from the inclusion of the determinant of the Ricci tensor in the action, thereby defining a new invariant scalar built from the complete Ricci tensor instead of only its trace. In \cite{Azri:2021det}, the authors derived the extended gravitational field equations of the Ricci-Determinant gravity (RD gravity) theory by independently varying the total action with respect to the metric and the affine connection. Subsequently, the authors constructed a stellar model, specifically a neutron star (NS) configuration, and derived the corresponding mass–radius relation by incorporating the effective contributions of the Ricci determinant within the Palatini formulation of gravity.\\

NSs are among the most abundant compact objects in galaxies throughout the Universe. They are observationally confirmed remnants formed from massive stars through a nonequilibrium gravitational collapse at the end of stellar evolution. Many theoretical studies suggest that the core of an NS may contain exotic constituents \cite{Bombaci:1997dd}, such as hyperons, kaon condensates, or a deconfined phase of strange matter. Other models propose the existence of hybrid stars composed of hadronic matter admixed with quark matter, possibly featuring a central core made entirely of deconfined quarks \cite{Glendenning:1995}. 
A quark star (QS) may be formed when the central density of a NS becomes sufficiently high to deconfine neutrons into their constituent quarks, leading to a phase transition to quark matter. Furthermore, QSs may form through various astrophysical processes, such as core collapse following a supernova explosion \cite{Dai:1995}, during which ordinary nuclear matter undergoes a phase transition into deconfined quark matter in the stellar core \cite{Cheng:1998wj}. In order to analyze how matter inside a compact star resists gravitational collapse at extreme densities and ultimately determines its structure and stability, the  EoS plays a crucial role. The EoS represents the fundamental relation between pressure and energy density of the matter distribution. For QSs, there are mainly two types of EoSs: the interacting quark matter EoS and the non-interacting quark matter EoS. Most analyses of QSs are still performed within the framework of the MIT bag model \cite{Chodos:1974je,Chodos:1974je1,Peshier:2000hx,Farhi:1984qu}. In this model, quarks confined inside the bag are treated as a free relativistic Fermi gas, while the bag constant provides an effective mechanism for quark confinement. This approach offers a simple description of a non-interacting quark matter EoS given by
\begin{equation}
p = \frac{1}{3}\left(\rho - 4B_{\mathrm{bag}}\right),
\end{equation}
where $p$ denotes the pressure, $\rho$ represents the energy density, and $B_{\mathrm{bag}}$ is the bag constant. However, according to the current understanding of quantum chromodynamics (QCD), QSs are far from being simple objects whose properties depend solely on the bag constant $B_{\mathrm{bag}}$ at high densities. A description based exclusively on this constant is insufficient, as it does not adequately account for the chiral symmetry properties inherent in QCD. Thus, several authors have proposed modified models such as three-flavor quark matter, called by the color-flavor-locked (CFL) phase of matter, which is believed to represent the true ground state of QCD at asymptotically high densities \cite{Alford:1998mk,Alford:1998mk1}. Physicists also propose several models that incorporate second and fourth order QCD corrections, aiming to provide a phenomenological account of quark confinement \cite{Flores:2017nonradial}. As a result, numerous models suggest that QSs composed of interacting quark EoS may exist at ultra-high densities \cite{Alford:2005js,Asbell:2017}. Details will be given in a later subsection. \\

Modern astrophysical observations have discovered many compact objects, improving our understanding of extremely dense gravitational systems. Some recently observed objects show properties that do not fully agree with standard GR or with simple NS EoS. In Refs.~\cite{Zhang:2021unified,Zhang:2021stellar}, the authors showed that the secondary component of the merger GW190814 could be a QS described by an interacting EoS. Additionally, several theoretical studies \cite{Holdom:2018hfd,Miao:2021xuq,Oikonomou:2023cflqs,Lopes:2022nature} have reported promising results concerning the interaction effects in different EoS and various quark matter phases. In Ref.~\cite{BecerraVergara:2019wgj}, it was shown that enhanced quark interactions allow QSs to sustain larger maximum masses, whereas weakly interacting quark matter produces results consistent with the MIT bag model. Although quark star solutions in the framework of 4D EGB gravity, with various EoSs, have been extensively investigated, they have produced several novel and significant results \cite{Gammon:2024quark,Banerjee:2021strange,Banerjee:2020xxx,Pretel:2022electrically,Rincon:2023anisotropic}. Alongside the EoS, novel insights into the physical properties of compact stars can be obtained by incorporating the effects of modified theories of gravity. For example, within the framework of Hořava–Lifshitz (HL) gravity, the values of the theory’s parameters may allow the existence of massive compact stars without requiring collapse into black holes \cite{Kim:2021neutronHL,Das:2025xxx}. Similar outcomes have also been reported in recently proposed extended gravity models, such as theories constructed from the determinant of the energy–momentum tensor \cite{Azri:2023detT,Das:2025possible}. Furthermore, in Ref.~\cite{Mota:2024xxx}, the authors demonstrated that gravitational modifications can significantly influence macroscopic properties of NSs, including their mass and radius.\\

Based on the present understanding of NSs and their internal composition, the possibility of more exotic compact objects cannot be excluded. Observational evidence suggests that such compact objects exhibit isotropic and homogeneous characteristics. Among the observable properties of a compact star are its mass, radius, and gravitational redshift. In particular, the mass and radius are of primary importance, as these quantities can be directly inferred from various astronomical observations. Their measured values provide crucial information about the global structure of the star and, consequently, offer deeper insights into the EoS governing its microscopic interior. Furthermore, recent advancements in observational technology have enabled stringent tests of GR and its extensions in the strong-field regime. Currently, investigations of the extreme physics governing NSs commonly incorporate the mass–radius constraints obtained from NICER observations \cite{Miller:2021c}, XMM-Newton data, and gravitational-wave measurements from the LIGO–VIRGO interferometric detectors \cite{Abbott:2017vtc}. Therefore, we are now in an era where advanced observational techniques enable us to gain deeper insights into compact stars and potential modifications of GR.\\

In the present study, we focus on a newly developed extension of GR, namely RD gravity. We consider a specific model in which the gravitational extension is characterized solely by the square root of the determinant of the Ricci tensor, which coincides with the well-known Eddington-inspired action \cite{Azri:2021det}. In this framework, we investigate isotropic and homogeneous QSs described by an interacting quark EoS under the effects of RD gravity. We derive and analyze the corresponding stellar structure equations for the functional form $f(\boldsymbol{D})=\lambda_{Edd}\sqrt{\boldsymbol{D}}$, where $\lambda_{Edd}$ is a dimensionless constant. Our primary motivation is to present a comparative analysis of the macroscopic properties, such as mass and radius of QSs in RD gravity, relative to those predicted by GR.

The organization of this paper is as follows: after this introductory section, we present a short recap of RD gravity in Section~\ref{sec2}. Next, the modified TOV equation, along with some other relevant equations that describe the structural framework of a static, spherically symmetric stellar object, are presented in Section~\ref{sec3}. In Section~\ref{sec4}, we review the details of the interacting quark EoS. After solving the TOV equations numerically, a detailed analysis of the physical properties, particularly the mass–radius relationship in RD gravity, is presented in Sections~\ref{sec5} and \ref{sec6}. Finally, in Section~\ref{sec7}, we summarize and discuss our key findings.

\section{Brief Introduction of RD Gravity}\label{sec2}

\noindent The Ricci-Determinant (RD) gravity is defined by the action: \cite{Azri:2021det}
\begin{eqnarray}\label{1}
\mathcal{S}=\int\sqrt{|\textbf{det.}g|}\Bigg[\frac{\mathcal{M}_{\text{pl}}^{2}}{2}\Big(g^{\mu\nu}\mathcal{R}_{\mu\nu}(\Gamma)\nonumber\\
-2\Lambda\Big)+\mathcal{L}_{M}[g]\Bigg]d^{4}x
+\int \sqrt{|\textbf{det.}g|}f(\boldsymbol{D})d^{4}x.
\end{eqnarray}
Here, $g$ is a Lorentzian metric, \textbf{det.} denotes the determinant, $\mathcal{L}_{M}[g]$ is the matter Lagrangian density independent of the connection $\Gamma$, $\mathcal{M}_{\mathrm{pl}}$ denotes the Planck mass, and $\Lambda$ is a constant. The function $f(\boldsymbol{D})$ is an arbitrary function of the scalar $\boldsymbol{D}$, defined as the ratio of two determinants, namely,
\begin{eqnarray}\label{2}
\boldsymbol{D}=\frac{|\textbf{det.}\mathcal{R}|}{|\textbf{det.}g|}.
\end{eqnarray}
By varying the action (\ref{1}) with respect to the metric tensor $g_{\mu\nu}$, we obtain the generalized Einstein field equations
\begin{eqnarray}\label{3}
\mathcal{R}_{\mu\nu}(\Gamma)=\Lambda g_{\mu\nu} + k\left[\mathcal{T}^{M}_{\mu\nu}- \frac{1}{2}g_{\mu\nu}g^{\alpha\beta}\mathcal{T}^{M}_{\alpha\beta}\right]\nonumber\\
+k\left(2\boldsymbol{D}f'(\boldsymbol{D})-f(\boldsymbol{D})\right)g_{\mu\nu},
\end{eqnarray}
where $\mathcal{T}^{M}_{\mu\nu}=g_{\mu\nu}\mathcal{L}_{M}-2\frac{\delta\mathcal{L}_{M}}{\delta g^{\mu\nu}}$ is the standard energy-momentum tensor of matter $k=1/\mathcal{M}_{pl}^{2}$ and $f'(\boldsymbol{D})=\frac{df}{d\boldsymbol{D}}$.\\
The second field equation is obtained from variation with respect to the connection, known as the dynamical equation, which leads to
\begin{eqnarray}\label{4}
\nabla_{\alpha}\left[\sqrt{|\textbf{det.}g|}g^{\mu\nu}+\frac{2\boldsymbol{D}f'(\boldsymbol{D})}{\mathcal{M}_{\text{pl}}^{2}}\sqrt{|\textbf{det.}g|}\left(\mathcal{R}^{-1}\right)^{\mu\nu}\right]=0.
\end{eqnarray}
We observe that the field equations derived from the action~(\ref{1}) deviate from the standard Palatini gravitational equations due to the presence of the function $f(\boldsymbol{D})$ and its derivative. As a consequence, the analytical solutions (\ref{3}) and~(\ref{4}) become highly nonlinear and mathematically intricate. Therefore, in the present work, we restrict our attention to a simpler yet physically interesting scenario in which Eq.~(\ref{3}) remains unaffected by the $f(\boldsymbol{D})$ term. In this case, the contribution from the Ricci-determinant term enters solely through Eq.~(\ref{4}).
\begin{eqnarray}\label{5}
2\boldsymbol{D}f'(\boldsymbol{D})-f(\boldsymbol{D})=0.
\end{eqnarray}
And solving the above equation (\ref{5}), we get
\begin{eqnarray}\label{6}
f(\boldsymbol{D})=\lambda_{Edd}\sqrt{\boldsymbol{D}},
\end{eqnarray}
where $\lambda_{Edd}$ is a dimensionless constant.\\
Therefore, the Eqs.~(\ref{3}) and (\ref{4}) can be modified as
\begin{eqnarray}\label{6a}
\mathcal{R}_{\mu\nu}(\Gamma)=\Lambda g_{\mu\nu} + k\left[\mathcal{T}^{M}_{\mu\nu}- \frac{1}{2}g_{\mu\nu}g^{\alpha\beta}\mathcal{T}^{M}_{\alpha\beta}\right]
\end{eqnarray}
and
\begin{eqnarray}\label{6b}
\nabla_{\alpha}\left[\sqrt{|\textbf{det.}g|}g^{\mu\nu}+\frac{\lambda_{Edd}}{\mathcal{M}_{\text{pl}}^{2}}\sqrt{|\textbf{det.}\mathcal{R}|}\left(\mathcal{R}^{-1}\right)^{\mu\nu}\right]=0.
\end{eqnarray}

\section{Structural Equations of Stellar Objects}\label{sec3}

\noindent In this section, we discuss the stellar structure, which is assumed to be static and spherically symmetric; accordingly, the spacetime geometry is described by the following line element
\begin{eqnarray}\label{10}
ds^{2}=-e^{2\mathcal{W}(r)}dt^{2}+e^{2\mathcal{H}(r)}dr^{2} + r^{2}\left(d\theta^{2} + \sin^{2}\theta d\phi^{2}\right),
\end{eqnarray}
where $W(r)$ and $H(r)$ are two arbitrary functions of radial coordinate $r$. \\
Next, we consider the interior matter distribution in an isotropic fluid mode, which is defined by the following equation:
\begin{eqnarray}\label{11}
\mathcal{T}_{\alpha\beta}=(\rho + p)\mathcal{U}_{\alpha}\mathcal{U}_{\beta} + pg_{\alpha\beta},
\end{eqnarray}
where $\rho$ and $p$ are the respective representations of energy density and pressure of the fluid. $\mathcal{U}^{\alpha}$ are 4-velocity vectors of the fluid, defined as $\mathcal{U}^{\alpha}=\left(e^{-\mathcal{W}}, 0, 0, 0\right)$.\\

Since Eq.~(\ref{6b}) involves the Ricci curvature, it is generally nonlinear in the affine connection $\Gamma$, making an explicit solution in terms of the metric potentially complicated. However, Eq.~(\ref{6a}) implies that the Ricci tensor can ultimately be expressed in terms of the matter fields. Consequently, Eq.~(\ref{6b}) becomes linear, as the connection appears only through the covariant derivative. To obtain an analytical solution, we therefore follow the standard procedure adopted in generalized Palatini theories by introducing an auxiliary tensor $h_{\alpha\beta}(x)$. The details of this approach have been discussed in Ref.~\cite{Azri:2021det}. Here, we directly employ the resulting auxiliary tensor, which takes the following form: \cite{Azri:2021det}
\begin{eqnarray}
h_{\alpha\beta}=-b\sqrt{(1+a)(1+a-b)^{-1}}\mathcal{U}_{\alpha}\mathcal{U}_{\beta}\nonumber\\
+\sqrt{(1+a-b)(1+a)}g_{\alpha\beta}
\end{eqnarray}
Here, the analytical expressions of $a$ and $b$ can be given as
\begin{eqnarray}
a=\frac{\lambda_{Ebd}}{2\mathcal{M}_{Pl}^{4}}\sqrt{(\rho-p)(\rho +3p)},
\end{eqnarray}
\begin{eqnarray}
b=\frac{\lambda_{Ebd}}{\mathcal{M}_{Pl}^{4}}(\rho +p)\sqrt{\frac{\rho-p}{\rho +3p}}=2a\left[\frac{\rho+p}{\rho+3p}\right].
\end{eqnarray}
Now, in terms of the tensor $h_{\alpha\beta}$, the gravitational field equation (\ref{6a}) can be rewritten in the following form
\begin{eqnarray}
\mathcal{R}_{\alpha}^{\beta}(h)=k\left[\mathcal{T}_{\alpha}^{\beta}-\frac{1}{2}\delta_{\alpha}^{\beta}\mathcal{T}\right]
\end{eqnarray}
Here, our primary motivation is to explore stellar structures, i.e., bounded systems in which gravity is governed predominantly by local mass–energy rather than by cosmic expansion or dark energy; hence, we have considered $\Lambda=0$.\\ 
The stellar structure corresponding to the metric~(\ref{10}) is described by two fundamental equations, adopted from Ref.~\cite{Azri:2021det}, which are given below:
\begin{eqnarray}
\frac{d\Psi}{dr}=\frac{k(\rho+p)r^{2}}{2(r-2m)}- \frac{k(\rho+p)r^{3}}{4(r-2m)}\left(a' - b'/2\right)\nonumber\\
-\frac{k(\rho +3p)r^{2}}{4(r-2m)}b-\frac{b'}{2}+r/2\left(a''-b''/2\right),
\end{eqnarray}
where $m(r)$ represents the total mass within the sphere of radius $r$.\\
\begin{eqnarray}
e^{2\mathcal{H}(r)}=\left(1-\frac{2m(r)}{r}\right)^{-1}~~~~,~~~~\Psi(r)=\mathcal{W}+\mathcal{H},
\end{eqnarray}
With this, the mass continuity equation is rewritten as:
\begin{eqnarray}
\frac{dm}{dr}=\frac{k\rho r^{2}}{2}+\left(r-\frac{3m}{2}-\frac{k\rho r^{3}}{4}\right)\left(a'-\frac{b'}{2}\right)\nonumber\\
+\frac{r^{2}}{2}\left(1-\frac{2m}{r}\right)\left(a''-\frac{b''}{2}\right)-\frac{k(\rho +3p)r^{2}}{8}b.
\label{eq:dmdr_mod}
\end{eqnarray}
Here, the expressions of first and second order derivatives of  $a$ and $b$ are given by
\begin{eqnarray}
a'=\frac{\lambda_{Ebd}}{4a\mathcal{M}_{Pl}^{8}}\left[\rho\rho'-3pp'+\rho p'+p\rho'\right]
\end{eqnarray}
\begin{eqnarray}
a''=\frac{\lambda_{Ebd}}{\mathcal{M}_{Pl}^{4}}\Big[(\rho+p)\rho''+(\rho-3p)p''+\rho'^{2}+2p'\rho'\nonumber\\
-3p'^{2}\Big]-a'^{2}
\end{eqnarray}
\begin{eqnarray}
b'=2a'\left[\frac{\rho+p}{\rho+3p}\right]+4a\left[\frac{p\rho'-\rho p'}{(\rho+3p)^{2}}\right]
\end{eqnarray}
\begin{eqnarray}
b''=2a''\left[\frac{\rho+p}{\rho+3p}\right]+8a'\left[\frac{p \rho'-\rho p'}{(\rho+3p)^{2}}\right]\nonumber\\
-8a\left[\frac{(p \rho'-\rho p')(\rho'+3p')}{(\rho+3p)^{2}}\right]+4a\left[\frac{p \rho''-p''\rho}{(\rho+3p)^{2}}\right]
\end{eqnarray}
The modified TOV equation i.e., the conservation equation $\nabla_{\mu}\mathcal{T}^{\mu\nu}=0$ reads as
\begin{eqnarray}
\frac{dp}{dr}=-(\rho +p)\mathcal{W}'\nonumber\\
=-(\rho +p)\left[\Psi'-\left(1-\frac{2m}{r}\right)^{-1}\frac{m'}{r}\right]\nonumber\\
-(\rho+p)\left(1-\frac{2m}{r}\right)^{-1}\frac{m}{r^{2}}
\end{eqnarray}
\begin{eqnarray}
\frac{dp}{dr}=-\frac{(\rho+p)}{r(r-2m)}\left(m+\frac{k~p~ r^{3}}{2}\right)+\frac{(\rho+p)}{2(r-2m)}\nonumber\\
\left(m+\frac{k~p~ r^{3}}{2}\right)\left(a'-\frac{b'}{2}\right)+(\rho +p)a'\nonumber\\
+\frac{k(\rho+p)(\rho+3p)r^{2}}{8(r-2m)}b
\label{eq:dpdr_mod}
\end{eqnarray}

\section{Interacting Quark Matter EOS}\label{sec4}

According to the Bodmer–Witten–Terazawa hypothesis \cite{Bodmer:1971we,Witten:1984rs,Terazawa:1979}, quark matter composed of nearly equal fractions of $u$, $d$, and $s$ quarks, known as strange quark matter (SQM), may constitute the absolute ground state of baryonic matter at zero temperature and pressure. Nevertheless, a recent investigation \cite{Holdom:2017iqi} indicates that $u$, $d$ quark matter (udQM) may generally be more stable than both SQM and conventional nuclear matter for sufficiently large baryon numbers extending beyond the periodic table. Interacting quark matter (IQM) incorporates interquark effects arising from strong interactions, including perturbative quantum chromodynamics (pQCD) corrections and color superconductivity \cite{Farhi:1984qu,Fraga:2001id,Fraga:2013qra,Alford:1998mk,Rajagopal:2000ff,Lugones:2002iz}. Recently, it has been shown that the EoS of IQM can be expressed in a simple unified form, and the resulting interacting quark stars (IQSs) are capable of satisfying various astrophysical constraints \cite{Zhang:2020qnw}. This development has motivated a series of follow-up studies \cite{Zhang:2021stellar,Zhang:2021xxx,Blaschke:2022crossover,Blaschke:2022,Gammon:2023xxx,Tangphati:2023,Yuan:2023tweflavor,Das:2024iqs,Das:2024cflstrangestars} employing this IQM EoS. In realistic stellar configurations, anisotropic effects should also be taken into account in QS, which have attracted considerable attention in recent years \cite{Komathiraj:2007fw,Harko:2002,Deb:2016nuv,Das:2023iqs,Lopes:2023xxx}. Very recently, the influence of anisotropy on the stellar structure of IQSs has been systematically investigated \cite{Pretel:2023xxx}.\\
According to the Refs.~\cite{Asbell:2017vdd,BecerraVergara:2019yjj}, the interacting quark EoS is given by
\begin{eqnarray}\label{EOS1}
p=\frac{1}{3}(\rho-4B_{\mathrm{bag}})-\frac{m_{s}^{2}}{3\pi}\sqrt{\frac{\rho-B_{\mathrm{bag}}}{a_{4}}}\nonumber\\
+\frac{m_{s}^{4}}{12\pi^{2}}\left[1-\frac{1}{a_{4}}+3\ln\left(\frac{8\pi}{3m_{s}^{2}}\sqrt{\frac{\rho-B_{\mathrm{bag}}}{a_{4}}}\right)\right] .
\end{eqnarray}
Here $\rho$ defines the energy density of homogeneously distributed quark matter (also to $\mathcal{O}(m_{s}^{4})$ in the Bag model) and $p$ is the pressure. Based on a review of the existing literature, we fix the strange quark mass to $m_{s}=100~\mathrm{MeV}$ \cite{Beringer:2012}. The commonly accepted range of the Bag constant $B_{bag}$ is $57~\mathrm{MeV/fm^{3}} \leq B_{bag} \leq 92~\mathrm{MeV/fm^{3}}$ \cite{Burgio:2018,Blaschke:2018}. The parameter $a_{4}$ comes from QCD corrections on the pressure of the quark-free Fermi sea, varying within $0< a_{4} \leq 1$; this parameter is related to the maximum mass of the star with approximate value $2 M_{\odot}$ at $a_{4} \approx 0.7$ \cite{Fraga:2000ew}.\\
Here we have used a unit conversion $1 \, \textrm{fm} = 197.3 \, \textrm{MeV}$ and hence $\textrm{MeV}^{4} = 197.3^{-3} \, \textrm{MeV}/\textrm{fm}^{3}$.\\
The above Eq.~(\ref{EOS1}) becomes
\begin{eqnarray}\label{eq:eosquark}
p=\frac{1}{3}(\rho - 4B_{\mathrm{bag}}) - \frac{1}{3\pi}\sqrt{\frac{1}{197.3^{3}}}m_{s}^{2}\sqrt{\frac{\rho - B_{\mathrm{bag}}}{a_{4}}}+\frac{1}{12\pi ^{2}}\nonumber\\
\frac{m_{s}^{4}}{197.3^{3}} \left[1- \frac{1}{a_{4}} + 3\ln\left(\frac{8\pi}{3m_{s}^{2}}\sqrt{197.3^{3}}\sqrt{\frac{\rho-B_{\mathrm{bag}}}{a_{4}}}\right)\right].
\end{eqnarray}
\begin{figure}
    \centering
    \includegraphics[width=.9\linewidth]{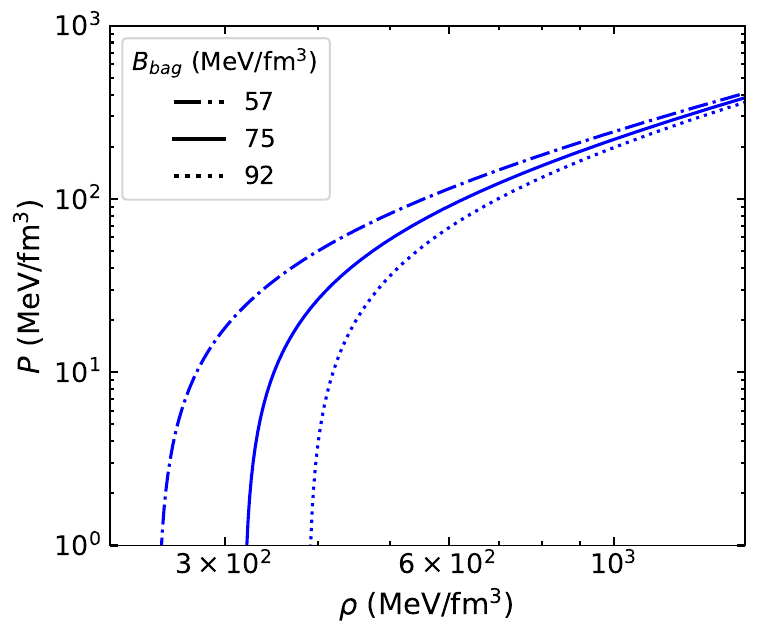}
    \caption{Interacting quark matter EOS [Eq.~(\ref{EOS1})] for different values of the bag constant. We select a central pressure of $P = 400 \, \mathrm{MeV/fm^{3}}$ corresponding to configurations near the maximum mass limit.}
    \label{fig:EOS(Prho)}
\end{figure}

\section{Numerical Methods and Analysis of the Mass–Radius Relation}\label{sec5}

In this section, the structure of quark stars composed of interacting quark matter is investigated within the framework of Ricci-Determinant Gravity. The modified stellar structure equations [Eqs.~ (\ref{eq:dmdr_mod}) and (\ref{eq:dpdr_mod})] are solved together with the quark matter EOS, Eq.~(\ref{eq:eosquark}). The stiffness of the EOS is regulated by the Bag constant 
$B_{bag}$, as illustrated in Fig.~\ref{fig:EOS(Prho)}. This dependence is exploited to constrain the parameter space of the theory by analyzing the impact of the coupling parameter $\lambda$ on macroscopic stellar properties, particularly the mass–radius relation, for different values of $B_{bag}$.

We adopt the perturbative approach proposed by \cite{Azri:2021det}, originally applied to hadronic objects with stiff EOSs (such as MPA1). In this formalism, the higher-order terms that arise in the modified metric factors are treated as perturbations on General Relativity (GR) in a way that the conservation of mass, Eq.~(\ref{eq:dpdr_mod}), and the modified TOV, Eq. (\ref{eq:dmdr_mod}), can be rewritten as:
\begin{eqnarray}
\frac{dm}{dr}= G_{GR} + \delta G,
\label{eq:dmdr_perturb}
\end{eqnarray}	
and
\begin{eqnarray}
\frac{dp}{dr}=F_{GR} + \delta F,
\label{eq:dpdr_perturb}
\end{eqnarray}
where
\begin{eqnarray}
G_{GR}= \frac{k\rho r^{2}}{2},
\end{eqnarray}
and
\begin{eqnarray}
F_{GR}= -\frac{(\rho+p)}{r(r-2m)}\left(m+\frac{k~p~ r^{3}}{2}\right).
\end{eqnarray}

For numerical integration, we define a dimensionless effective coupling parameter normalized by the Planck mass scale ($M_{Pl} = 1.22 \times 10^{19}$ GeV):
\begin{align}
    \lambda \equiv \frac{\lambda_{Edd}}{M_{Pl}^4},
\end{align}
where $\lambda_{Edd}$ is the parameter associated with the Eddington scale, acting as a limiting reference in the theory. We then integrate the equations from the center ($r=0$, $m=0$, where the pressure $P_c$ and density $\rho_c$ are maximum) to the surface ($P \to 0$).

\subsection{General RD Gravity effects}

\begin{figure*}[ht!]
    \centering
    \includegraphics[width=.9\linewidth]{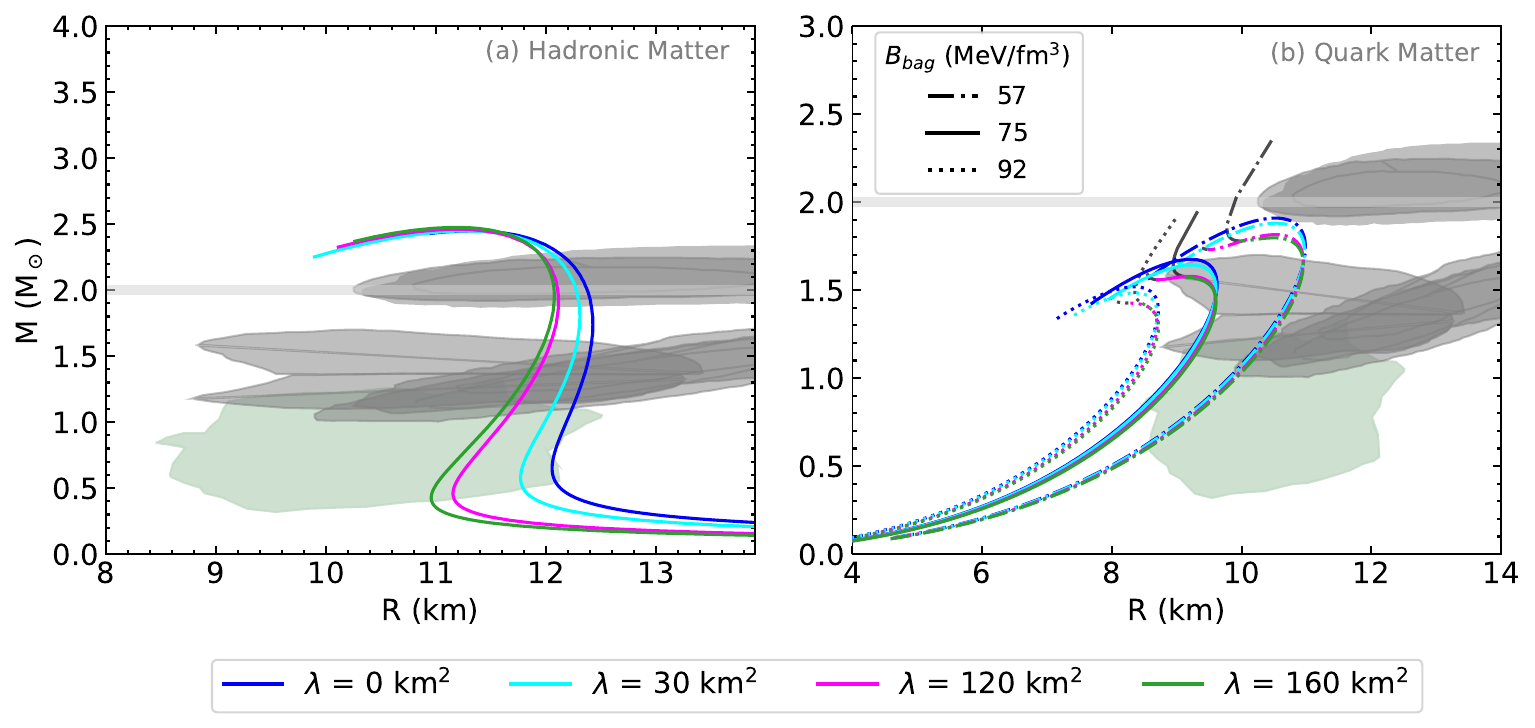}
    \caption{Mass–radius relations in the Ricci-determinant model governed by the geometrically effective parameter $\lambda$, where $\lambda = 0$ represents the ordinary case (GR). Panel (a) shows the hadronic case (EOS MPA1), with results taken from \cite{Azri:2021det}, while panel (b) presents our original results for compact stars composed of interacting quark matter with different bag constants. Black segments of the curves indicate RD gravity-dominant configurations. The observational constraints are shown in the background, highlighting the quark-star candidate HESS J1731 (green region).}
    \label{fig:MR}
\end{figure*}
Fig.~\ref{fig:MR} compares the hadronic scenario (panel (a), adapted from Ref.~\cite{Azri:2021det}) with our results for quark stars [panel (b)]. To enable a direct comparison, we consider the values $\lambda = 30$, $120$, and $160$, with $\lambda = 0$ corresponding to the GR case (blue curve). In the hadronic case, in agreement with previous results, increasing $\lambda$ leads to more compact objects, resulting in a reduced stellar radius for a fixed central density. At the same time, the maximum supported mass increases, indicating an enhanced gravitational support.

In contrast, our results for quark stars exhibit a qualitatively different behavior. Unlike the hadronic scenario, increasing $\lambda$ produces a slight increase in stellar radii, suggesting that the modified gravity effects act to reduce the effective compactness in order to maintain hydrostatic equilibrium. The corresponding gain in maximum mass is less pronounced than in the hadronic case and becomes noticeable only at high central densities, as indicated by the bounds of the black curves. Consequently, as the parameter $\lambda$ increases, the maximum mass shifts toward smaller values.

\begin{figure*}
    \centering
    \includegraphics[width=.9\linewidth]{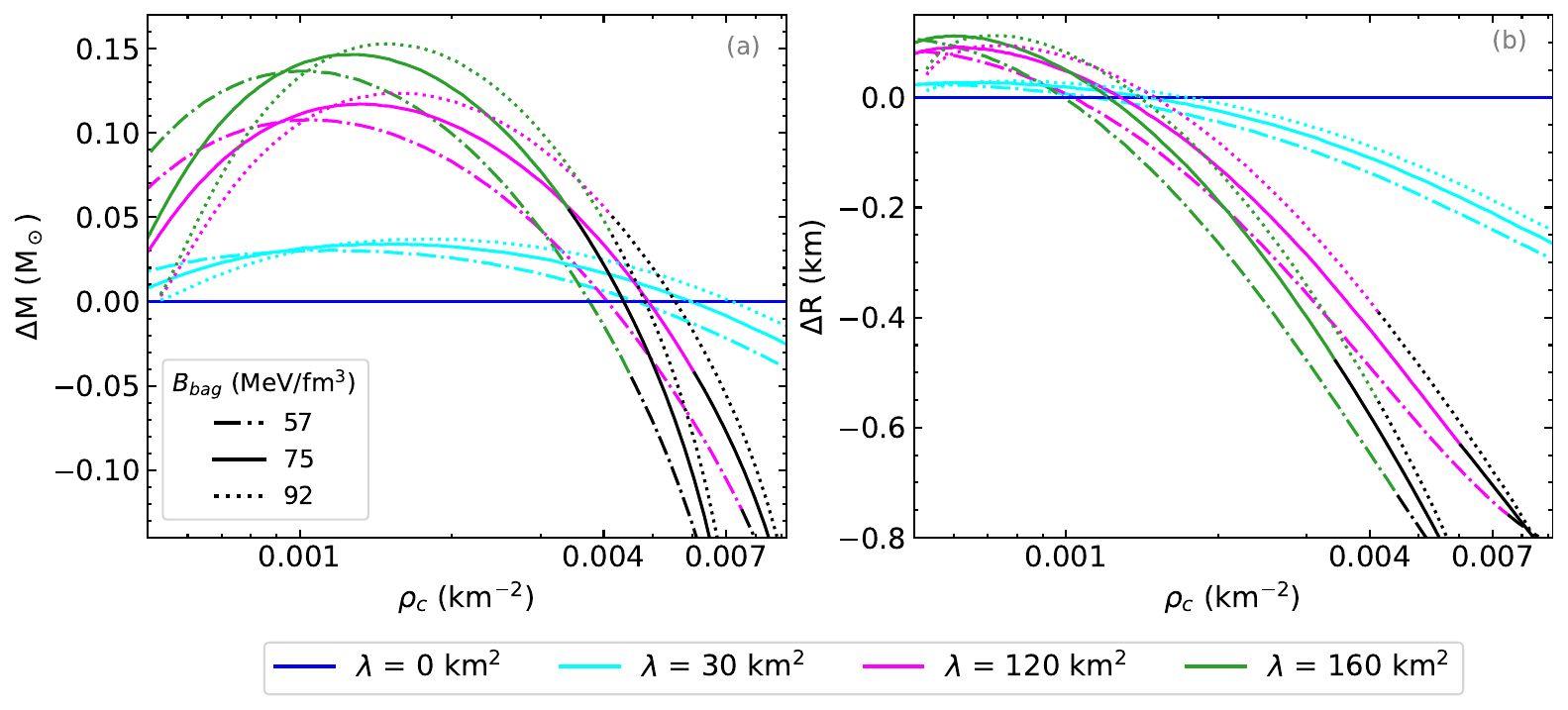}
    \caption{Mass and radius deviations from general relativity in the Ricci-determinant model for different values of $\lambda$. The deviations are defined as $\Delta \textrm{M} = \textrm{M}_{GR} - \textrm{M}_{RD}$ and $\Delta \textrm{R} = \textrm{R}_{GR} - \textrm{R}_{RD}$ shown for different bag constants $B_{bag}$ as functions of the central density. The radius [panel (b)] exhibits an upward deviation from the general-relativistic prediction ($\lambda = 0$; $\Delta \textrm{R} < 0$) already at low central densities, whereas the stellar mass [panel (a)] exceeds its GR counterpart ($\Delta \textrm{M} < 0$) only at high central densities. Black segments of the curves indicate RD gravity-dominant configuration.}
    \label{fig:deltaMR}
\end{figure*}

To better interpret the results, we present in Fig.~\ref{fig:deltaMR} the relative differences in mass and radius as functions of the central density $\rho_c$, defined as
\begin{align}
\Delta M &= M_{GR} - M_{RD}, \\
\Delta R &= R_{GR} - R_{RD}.
\end{align}
In panel (b), it is noted that $\Delta R$ assumes negative values, confirming that $R_{RD} > R_{GR}$. That is, Ricci-Determinant Gravity tends to expand the stellar radius, moving away from the relativistic case even at low densities ($\rho_c \gtrsim 0.001$ km$^{-2}$), an effect accentuated for more rigid EOSs ($B_{bag} = 57$ MeV/fm$^3$). In panel (a), mass variations become relevant only at high central densities ($\rho_c \gtrsim 0.004$ km$^{-2}$), corroborating that the impact on the maximum mass occurs predominantly in the strong-curvature regime. 
\begin{figure}[ht]
    \centering
    \includegraphics[width=0.9\linewidth]{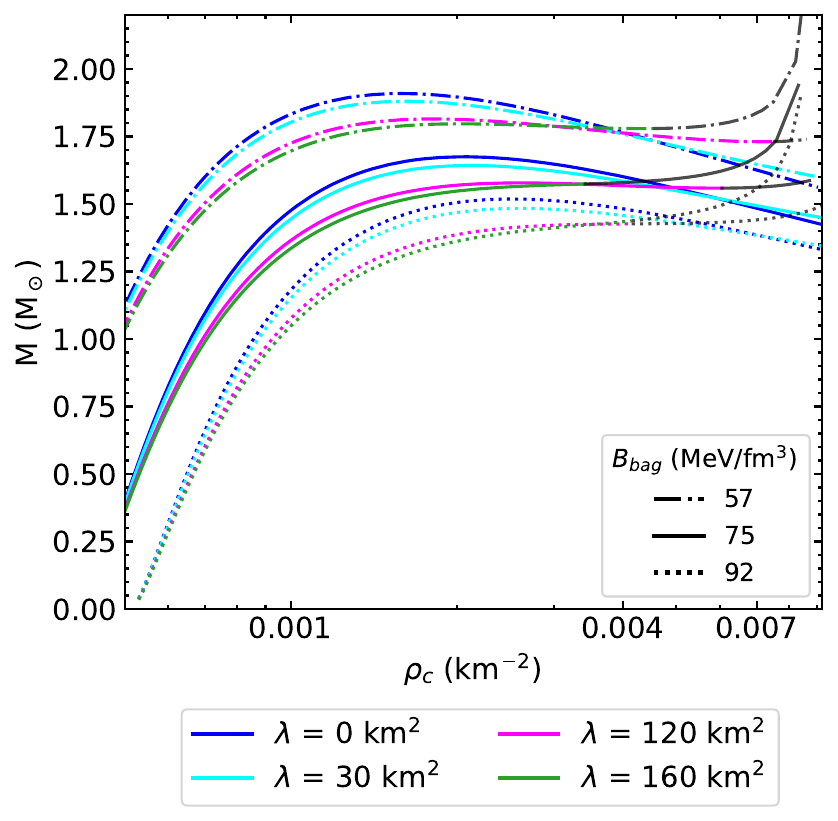}
    \caption{Total mass in Ricci-determinant gravity for stability configurations with different central densities. Black segments of the curves indicate RD gravity-dominant configuration and arise for central densities exceeding $\sim 0.004$ km${}^{-2}$ at $\lambda \approx 120$.}
    \label{fig:Mec-stability}
\end{figure}
Fig.~\ref{fig:Mec-stability} shows that, for central densities above this limit (to the left of the peak in the M–R diagram), the change in inclination reflects a modification of the stability properties inferred from the turning-point criterion. We observe that increasing the parameter $\lambda$ slightly shifts the instability point ($\partial M / \partial \rho_c = 0$), allowing for distinct maximum-mass configurations while maintaining the general topology of the stability transition.
\begin{figure*}
    \centering
    \includegraphics[width=.9\linewidth]{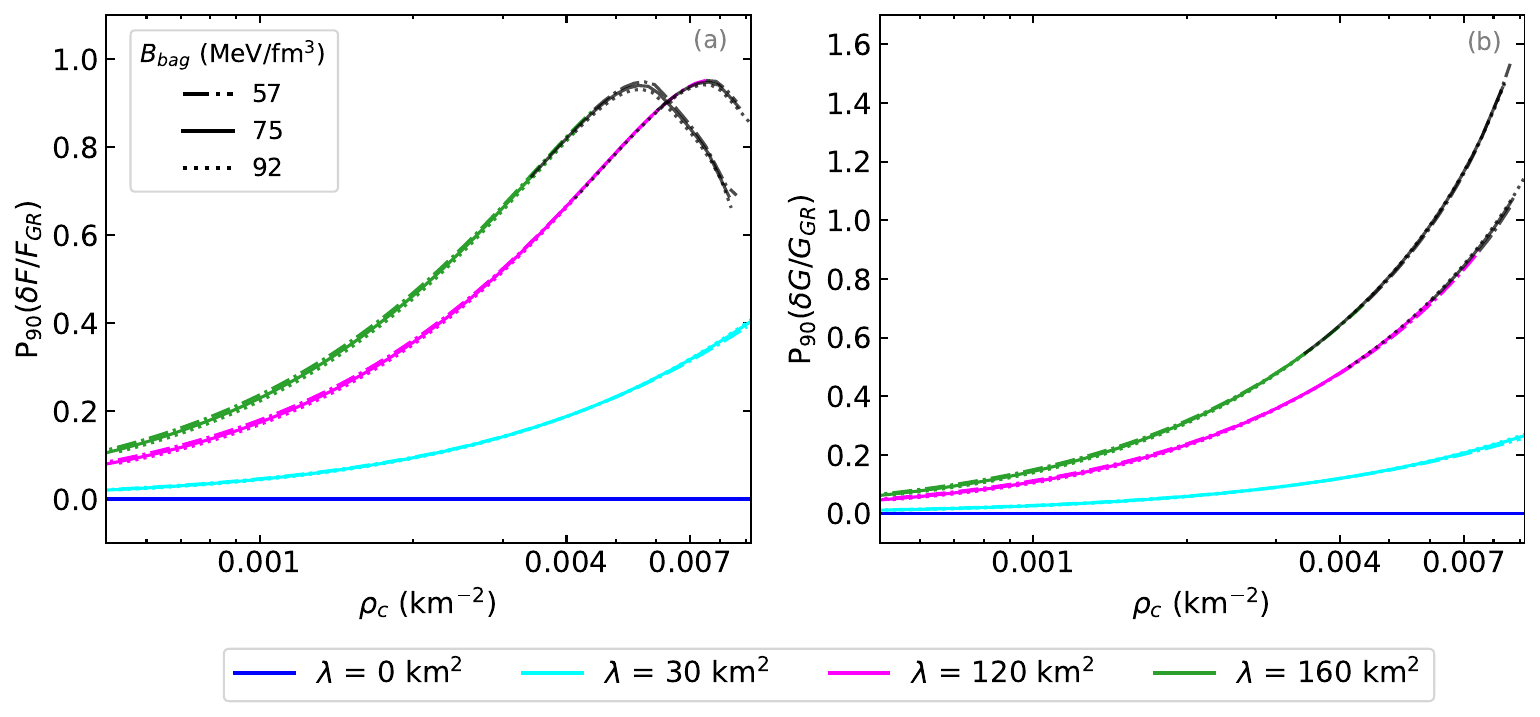}
    \caption{Relative contribution of the Ricci-determinant gravity corrections with respect to GR ($\lambda = 0$). In panel (a), $F$ represents the modification of the TOV equation, with $\delta$ denoting perturbative contributions, while in panel (b), $G$ denotes the modification of the mass continuity equation. The black segments of the mass-radius diagram denote regions where both $\delta F/F_{\textrm{GR}}$ and $\delta G/G_{\textrm{GR}}$ exceed $0.5$, and the transition to a steeper mass–radius relation occurs when $\delta F/F_{\text{GR}} \sim 1$ and $\delta G/G_{\textrm{GR}} > 1$. Owing to the strong radial dependence of these quantities, percentiles of their distributions (specifically the 90th percentile, P${}_{90}$) provide representative measures.}
    \label{fig:deltaFG}
\end{figure*}
Fig.~\ref{fig:deltaFG} clarifies the physical origin of these modifications. We analyze the 90th percentiles ($P_{90}$) of the radial distributions of the perturbative terms $\delta G/G_{GR}$ [Eq.(\ref{eq:dmdr_perturb})] and $\delta F/F_{GR}$ [Eq.(\ref{eq:dpdr_perturb})], which provide a statistically robust characterization of their maximal behavior along the stellar profile.

The nonlinear coupling between the corrections $\delta F$ and $\delta G$ results in a dominant relative modification of the hydrostatic equilibrium, while the effective mass is comparatively less affected at first. For instance, consider a fixed central energy density $\rho_c \sim 0.001 \, \textrm{km}^{-2}$, we find:
\begin{align}
 \lambda = 30 \, \textrm{km}^{2}: \delta F/{F_{\rm GR}} \sim 0.04,\, \delta G/{G_{\rm GR}} \sim 0.02; \\
\lambda = 120 \, \textrm{km}^{2}: \delta F/{F_{\rm GR}} \sim 0.17, \, \delta G/{G_{\rm GR}} \sim 0.10; \\
\lambda = 160 \, \textrm{km}^{2}: \delta F/{F_{\rm GR}} \sim 0.22, \, \delta G/{G_{\rm GR}} \sim 0.14,
\end{align}
indicating that the additional term in the hydrostatic equilibrium equation provides the leading contribution over a large range of central densities. As a consequence, the extra hydrostatic support favors the formation of quark stars with larger radii, whereas the effective mass remains self-regulated by the mass equation and does not exhibit a proportional increase.

This behavior is different in the hadronic case due to the specific properties of the equation of state. In quark stars, the equation of state implies that, at the stellar surface, the pressure vanishes while a finite and relatively large energy density is still present (see Fig.\ref{fig:EOS(Prho)}). As a result, the effects of RD gravity terms remain relevant throughout the stellar interior. In contrast, for a hadronic EOS, the energy density becomes very small at the surface, causing the corrections $\delta F$ and $\delta G$ to be significant mainly in the stellar core. Consequently, although the total effective mass is affected, the additional hydrostatic support is predominantly confined to the high-density core, and the pressure reaches zero at smaller radii, leading to more compact configurations.

The regions highlighted in black in the mass--radius diagram correspond to configurations in which the perturbative contributions exceed half of the general relativistic terms ($\delta F/F_{GR} > 0.5$ and $\delta G/G_{GR} > 0.5$). We find that the steeper change in the slope of the mass--radius curves coincides with the regime in which these corrections become dominant. In particular, the term $\delta G$, associated with the geometric effective mass, plays the leading role in modifying the equilibrium configurations and increasing the maximum mass. This behavior, however, signals the onset of a regime incompatible with the perturbative approximation, as the corrections from RD gravity become comparable to or larger than the general relativistic contributions. Since $\delta G/G_{GR}$ grows monotonically with the central density $\rho_c$, while $\delta F/F_{GR}$ increases only up to a maximum and subsequently decreases, the pressure--gravity balance is progressively shifted and can formally induce a recovery of $\partial M/\partial\rho_c > 0$, without implying the existence of a controlled additional equilibrium branch.

\subsection{Observational constraints}

The mass–radius diagram (Fig.~\ref{fig:MR}) also includes observational constraints from the pulsars PSR J0740+6620 and PSR J0030+0451, the merger event GW170817, and the quark-star candidate HESS J1731 (highlighted in green). We find that stiffer equations of state (EOS), corresponding to smaller values of the bag constant $B_{\mathrm{bag}}$, favor higher maximum masses and are therefore more compatible with the existence of massive pulsars. Regarding the coupling parameter $\lambda$, we observe that relatively small values ($\lambda \lesssim 30 \, \mathrm{km}^2$) are already sufficient to modify the compatibility with the $2 \, M_\odot$ constraint. For the candidate HESS J1731, the observational constraints are better satisfied for $B_{\mathrm{bag}} \approx 57 \, \mathrm{MeV/fm}^3$ across the entire range of $\lambda$ considered. In contrast, for $B_{\mathrm{bag}} \approx 75 \, \mathrm{MeV/fm}^3$, the constraints are only marginally satisfied, particularly for configurations with higher central densities and with respect to the inferred radius limits when $\lambda \gtrsim 120 \, \mathrm{km}^2$.

Our results indicate that noticeable macroscopic effects in quark stars emerge when the coupling parameter $\lambda$ is of the order of tens of $\mathrm{km}^2$. However, increasing $\lambda$ also enlarges the instability region at high central densities, thereby reducing the range of stable equilibrium configurations, as discussed previously. Consequently, in order to reconcile Ricci-Determinant Gravity with the existence of massive and stable compact objects, $\lambda$ must remain below values of the order of a few hundred $\mathrm{km}^2$ ($10 \lesssim \lambda \lesssim 100 \, \mathrm{km}^2$). This upper bound becomes more restrictive for stiffer EOSs, as indicated by the increase in the black segments in Fig.~\ref{fig:MR}.

\section{Structural Aspects of Quark Stars}\label{sec6}

To ensure completeness, we analyze several relevant physical characteristics within the interior of a fluid sphere in stable configurations.

\subsection{Stability Analysis and Corresponding Adiabatic Indices}
\begin{figure}[ht]
    \centering
    \includegraphics[width=0.85\linewidth]{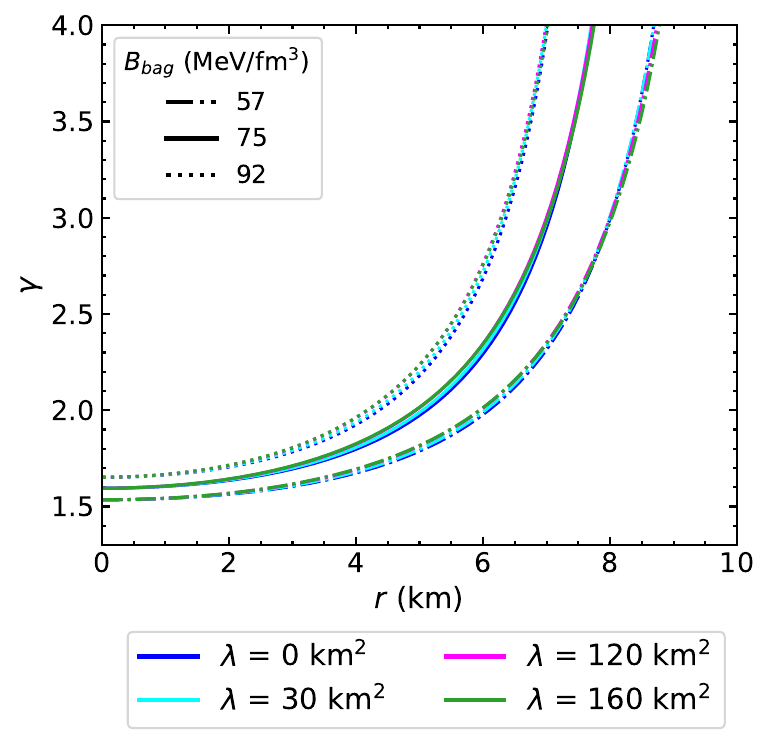}
    \caption{Behavior of the adiabatic index ($\gamma$) for objects with a central pressure of $P = 400 \, \mathrm{MeV/fm^{3}}$. The different curves correspond to varying values of the bag constant ($B_{\mathrm{bag}}$), with minor variations arising from RD gravity effects.}
    \label{fig:GAMMAr}
\end{figure}

Here, we examine the stability of the present QS stellar model by analyzing the behavior of the adiabatic index ($\gamma$) in the interior. This quantity plays a fundamental role in the instability criterion and is related to a thermodynamic parameter introduced by Chandrasekhar in 1964. The adiabatic index \cite{Chandrasekhar:1964zz,Chan:1993} can be defined as follows:
\begin{eqnarray}
\gamma=\left(1 + \frac{\rho}{p}\right)\left(\frac{dp}{d\rho}\right).
\end{eqnarray}
It is important to mention that, for the stability of compact relativistic objects, including white dwarfs, NSs, and supermassive stars, the values of $\gamma$ must exceed $4/3$ at every point within the interior of the configuration \cite{Glass:1983stability,Moustakidis:2017tlk}. We have illustrated the behavior of the adiabatic index, $\gamma$, as a function of the radial coordinate of the stellar object in Fig.~\ref{fig:GAMMAr} corresponding to different considered values of $B_{bag}$. It is evident from the figure that $\gamma > 4/3\sim 1.33$ throughout the stellar interior, indicating that the configuration remains stable against radial adiabatic infinitesimal perturbations. Moreover, larger values of $\gamma$ imply a stronger increase in pressure for a given increment in energy density, corresponding to a stiffer EoS. The effects of RD gravity are subtle over most of the objects’ interiors, but they become especially noticeable at intermediate densities.

\subsection{Compactness and Binding Energy}

\begin{figure}[ht]
    \centering
    \includegraphics[width=0.9\linewidth]{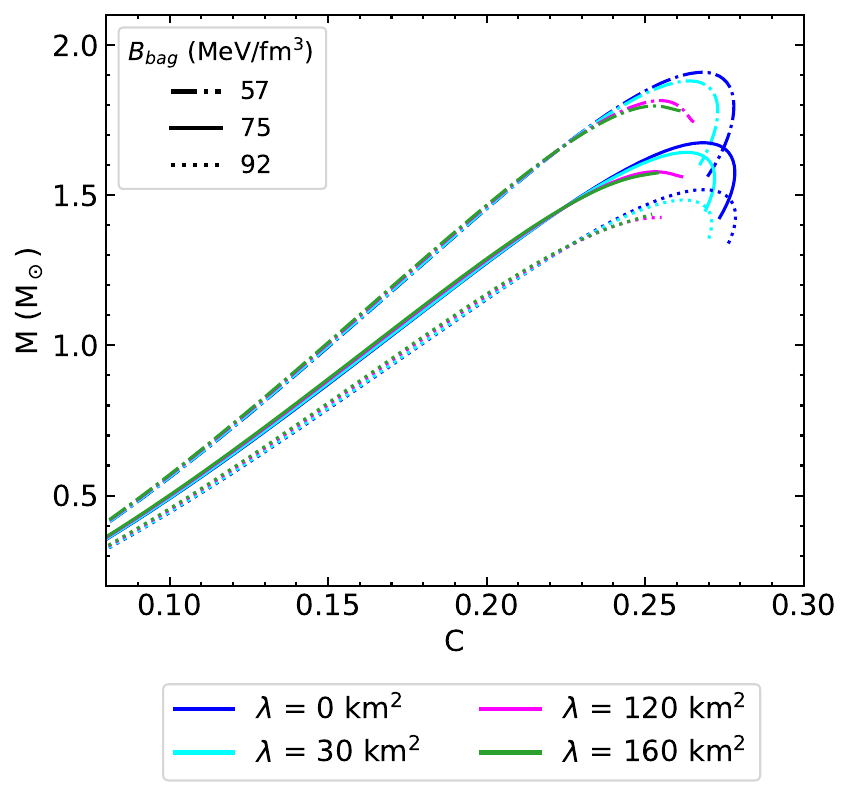}
    \caption{Mass-compactness relations for objects with different central densities. RD gravity effects become significant for massive objects.}
    \label{fig:CompacM}
\end{figure}

\begin{figure*}[ht]
    \centering
    \includegraphics[width=0.9\linewidth]{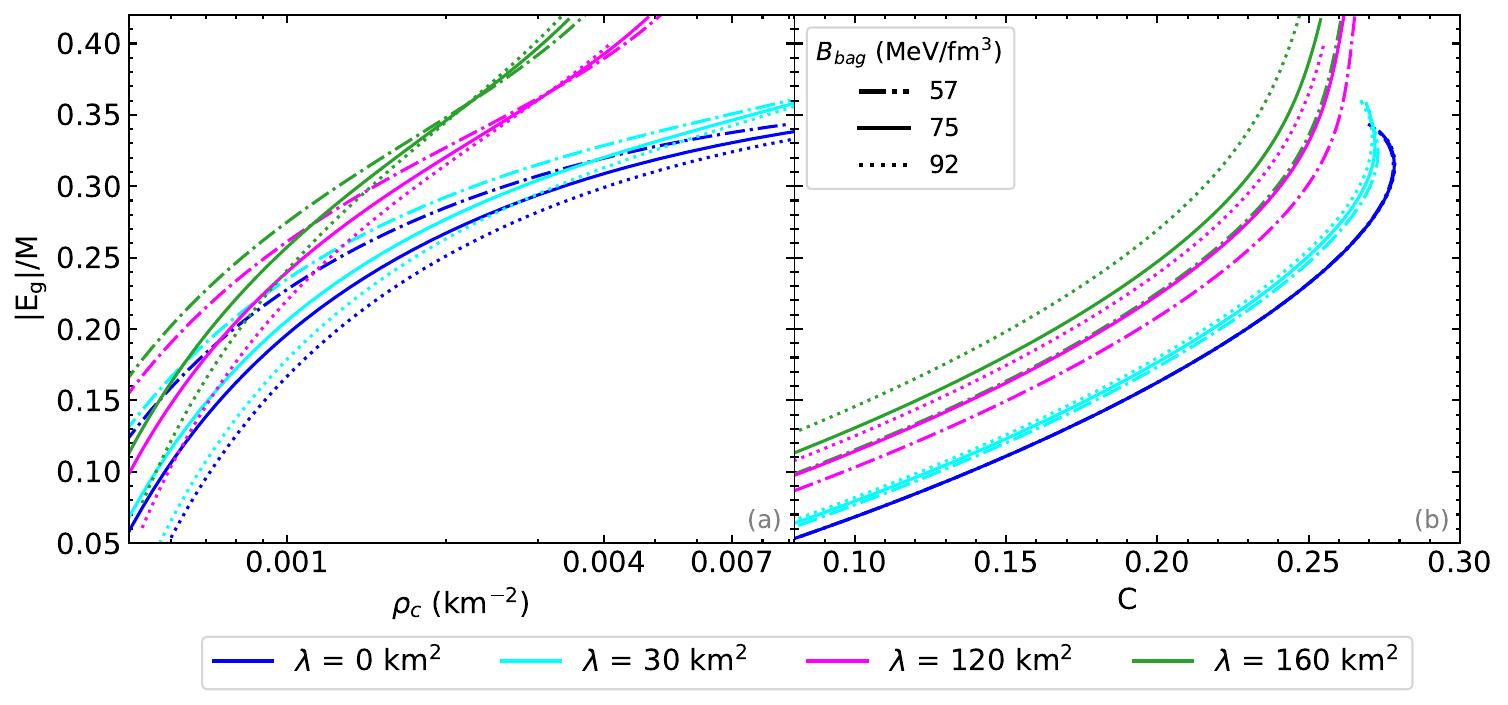}
    \caption{Analysis of the gravitational binding energy in RD gravity. Panel (a) shows its dependence on the central density, while panel (b) shows its dependence on the compactness.}
    \label{fig:CompacEg}
\end{figure*}

In compact relativistic stars, the compactness parameter plays a crucial role and is defined as:
\begin{eqnarray}\label{Com1}
C = \frac{GM}{Rc^2}.
\end{eqnarray}
For physically well-behaved compact stars, the compactness $C$ should not exceed $4/9$, as proposed by Buchdahl \cite{Buchdahl:1959zz}. For the present model, the $M-C$ relation is illustrated in Fig.~\ref{fig:CompacM} for different values of $B_{\mathrm{bag}}$. It can be observed that the compactness $C$ depends on the model parameter $\lambda$; however, it does not violate the Buchdahl condition for any of the considered variations. In all cases, we obtain $C<4/9$, indicating that the deformation introduced by Lorentz-symmetry breaking preserves the geometric and causal admissibility of the solutions.\\
The gravitational binding energy (GBE) is an important physical quantity that can provide insight into the internal structure of a NS \cite{Jiang:2019xwn}. Physically, it represents the energy required to disperse all the matter of a compact object against its own gravitational field and thus provides a measure of how strongly the stellar matter is gravitationally bound. Since the binding energy depends sensitively on the mass distribution and the equation of state of dense matter, it can offer useful insight into the internal composition and stability of compact stellar objects. In this context, it is also meaningful to evaluate the GBE for our QS stellar configurations. To proceed, we begin with the definition of the proper mass of a compact star, which is expressed as \cite{Bagchi:2011}
\begin{eqnarray}\label{eq:Eg}
\textrm{E}_\text{g} = \textrm{M} - \textrm{M}_\text{p}, 
\end{eqnarray}
where
\begin{eqnarray}\label{eq:Eg}
\textrm{M}_\text{P} = \int_{0}^{R}{4 \pi r^{2} \rho\,\left(1-\frac{2m(r)}{r}\right)^{-1} dr}. 
\end{eqnarray}
In the left panel (a), we present the behavior of the absolute value of the GBE as a function of $\rho_{c}$, while the right panel (b) of Fig.~\ref{fig:CompacEg} illustrates the variation of the absolute value of the GBE with respect to $C$. For every considered value of $B_{\mathrm{bag}}$, the absolute value of the GBE increases with the increasing central density $\rho_{c}$, and a similar behavior is observed with respect to the compactness $C$. Thus, within the framework of RD gravity, a more massive and comparatively more compact QS may be theoretically possible.
\begin{table*}[tb]
\centering
\caption{Maximum mass configurations for different values of $B_{bag}$.}
\label{tb:table1}
\renewcommand{\arraystretch}{1.2}

\begin{tabular}{ccccccc}
\hline\hline
\multicolumn{7}{c}{$B_{bag} = 57\,\mathrm{MeV/fm^3}$} \\
\hline
$\lambda$ (km${}^2$) & $\rho_c$ (MeV/fm${}^3$) & $\gamma_c$ & $M_{\max}(M_\odot)$ & $R$ (km) & $C_{max}$ & E${}_\text{g}$/M${}_{max}$ \\
\hline
0   & 1192 & 1.40 & 1.90 & 10.52 & 0.27 & 0.27 \\
30  & 1240 & 1.40 & 1.88 & 10.50 & 0.26 & 0.28 \\
120 & 1395 & 1.39 & 1.82 & 10.50 & 0.25 & 0.32 \\
160 & 1512 & 1.39 & 1.80 & 10.50 & 0.25 & 0.33 \\

\hline
\multicolumn{7}{c}{$B_{bag} = 75\,\mathrm{MeV/fm^3}$} \\
\hline
$\lambda$ (km${}^2$) & $\rho_c $ (MeV/fm${}^3$) & $\gamma_c$ & $M_{\max}(M_\odot)$ & $R$ (km) & $C_{max}$ & E${}_\text{g}$/M${}_{max}$ \\
\hline
0   & 1585 & 1.40 & 1.67 & 9.21 & 0.27 & 0.27 \\
30  & 1585 & 1.40 & 1.65 & 9.20 & 0.26 & 0.28 \\
120 & 2012 & 1.39 & 1.58 & 9.18 & 0.25 & 0.35 \\
160 & 2584 & 1.39 & 1.57 & 9.15 & 0.25 & 0.42 \\

\hline
\multicolumn{7}{c}{$B_{bag} = 92\,\mathrm{MeV/fm^3}$} \\
\hline
$\lambda$ (km${}^2$) & $\rho_c$ (MeV/fm${}^3$) & $\gamma_c$ & $M_{\max}(M_\odot)$ & $R$ (km) & $C_{max}$ & E${}_\text{g}$/M${}_{max}$\\
\hline
0   & 1852 & 1.40 & 1.52 & 8.37 & 0.27 & 0.27 \\
30  & 1925 & 1.40 & 1.48 & 8.37 & 0.26 & 0.28 \\
120 & 2545 & 1.39 & 1.42 & 8.36 & 0.25 & 0.37 \\
160 & 2545 & 1.39 & 1.42 & 8.48 & 0.25 & 0.42 \\

\hline\hline
\end{tabular}
\end{table*}

A summary of the discussed effects of RD gravity on interacting quark stars can be seen in \autoref{tb:table1}. We can highlight that among the quantities analyzed in this section, the gravitational binding energy is the most susceptible to modifications, leading to limiting configurations with increasing $\lambda$. For instance, $\lambda = 160$ reports E${}_\text{g}$/M${}_{max} > 0.40$, an extreme limit that appears in complex formulations for interacting quark matter in the context of relativity. On the other hand, a central adiabatic index ($\gamma_c$) is fixed around $1.40$ on maximum mass configurations, and the compactness stands out in a range of 0.25 to 0.27, showing a decrease with $\lambda$.

\section{Conclusions}\label{sec7}
In the present work, we investigate several fundamental properties of QSs composed of interacting quark matter described by a nonlinear EoS. For this purpose, we consider a recently developed extension of GR, namely RD gravity \cite{Azri:2021det}. Unlike GR, this theory is motivated by Eddington gravity as well as Eddington-inspired Born-Infeld gravity and is formulated in terms of the determinant of the Ricci tensor. In this framework, the Palatini action is generalized by introducing an arbitrary function $f(\boldsymbol{D})$ of the Ricci determinant, where $\boldsymbol{D}=\frac{|\textbf{det.}\mathcal{R}|}{|\textbf{det.}g|}$. 
We focus on a particular model arising from this general formulation, which corresponds to the addition of the well-known Eddington action involving the square root of the Ricci determinant; further details can be found in Ref.~\cite{Azri:2021det}. In particular, we analyze the mass-radius relation and the stability properties of the proposed QS configurations within the context of RD gravity. Since the modified TOV equations governing the stellar structure are highly nonlinear, we solve them numerically. By varying two model parameters, namely the gravity parameter $\lambda$ and the bag constant $B_{\mathrm{bag}}$ from the EoS, we perform a detailed graphical analysis and compare the deviations of RD gravity from the standard GR predictions.

We see that the effects of RD gravity on the properties of quark stars manifest in distinct ways in the general equations. The inclusion of $\delta F$ alters the hydrostatic equilibrium, favoring the formation of objects with larger radii. On the other hand, $\delta G$ promotes an increase in mass in objects with high central densities, although on a reduced scale. The validity of the perturbative approximation is compromised when $\delta F$ and $\delta G$ exceed 60\% of $F_{GR}$ and $G_{GR}$, respectively. Under these conditions, instabilities arise that prevent the formation of more massive configurations.

We observe direct effects of quark star structure as a result of perturbative physical mechanisms. These impacts become evident when $\lambda$ assumes values on the order of tens. Unlike the hadronic scenario, increasing this parameter raises both the radius and the mass, resulting in a less compact object. This increase also shifts the stability limit ($\partial M/\partial\rho_c = 0$), imposing restrictions on the formation of objects with high central densities, as observed at $\lambda = 160$, where $\rho_{\text{max}} \lesssim 0.04$.

It is clear that the rigidity of the EoS is a determining factor in defining the parameters of the theory. Taking into account their characteristics, it is possible to establish an upper limit of $\lambda = 100$ for $B_{\text{bag}} < 75$ MeV fm$^{-3}$, highlighting that the rigidity of the EoS is determinant for these observational aspects. In the case of $B_{\text{bag}} = 56$ MeV fm$^{-3}$, it is also possible to note that even a moderate value of $\lambda = 30$ already violates the maximum mass limit established for pulsars. 

The theoretical consistency of the model is supported by the analysis of the adiabatic index, which always remains greater than $\gamma = 4/3$, and the compactness, which remains within the causal limit. From a structural point of view, the gravitational binding energy (GBE) tends to increase with $\lambda$, reaching extreme regimes up to the stability threshold. Furthermore, we verified that the variation of the parameter $\lambda$ implies a break in the universality between compactness and GBE, as discussed in \cite{Jiang:2019xwn}.

Motivated by technological advancements in astronomical instrumentation, we have established new criteria for verifying the RD gravity theory, defining observable parameters in a quark star. However, the existence of these objects remains unconfirmed. Moreover, validating these hypotheses requires a more rigorous comparison with General Relativity across different astrophysical and cosmological environments. Therefore, further investigations are needed.\\

\textbf{Acknowledgements:} L.F.A. acknowledges the financial support from Coordenação de Aperfeiçoamento de Pessoal de Nível Superior (CAPES), Brazil.\\

\textbf{Data Availability Statement:} Data sharing is not applicable to this article as no datasets were used.

\newpage

\bibliographystyle{elsarticle-num}

\bibliography{ref}

\begin{thebibliography}{10}
\expandafter\ifx\csname url\endcsname\relax
  \def\url#1{\texttt{#1}}\fi
\expandafter\ifx\csname urlprefix\endcsname\relax\def\urlprefix{URL }\fi
\expandafter\ifx\csname href\endcsname\relax
  \def\href#1#2{#2} \def\path#1{#1}\fi

\bibitem{Planck:2018vyg}
P.~Collaboration, N.~Aghanim, et~al., {Planck 2018 results. VI. Cosmological parameters}, Astron. Astrophys. 641 (2020) A6.
\newblock \href {http://arxiv.org/abs/1807.06209} {\path{arXiv:1807.06209}}, \href {https://doi.org/10.1051/0004-6361/201833910} {\path{doi:10.1051/0004-6361/201833910}}.

\bibitem{Riess:1998cb}
A.~G. Riess, A.~V. Filippenko, et~al., {Observational Evidence from Supernovae for an Accelerating Universe and a Cosmological Constant}, Astron. J. 116 (1998) 1009--1038.
\newblock \href {http://arxiv.org/abs/astro-ph/9805201} {\path{arXiv:astro-ph/9805201}}, \href {https://doi.org/10.1086/300499} {\path{doi:10.1086/300499}}.

\bibitem{Ishak:2018his}
M.~Ishak, {Testing General Relativity in Cosmology}, Living Rev. Relativ. 22~(1) (2019) 1.
\newblock \href {http://arxiv.org/abs/1806.10122} {\path{arXiv:1806.10122}}, \href {https://doi.org/10.1007/s41114-018-0017-4} {\path{doi:10.1007/s41114-018-0017-4}}.

\bibitem{Nojiri:2017ncd}
S.~Nojiri, S.~D. Odintsov, V.~K. Oikonomou, {Modified gravity theories on a nutshell: Inflation, bounce and late-time evolution}, Phys. Rept. 692 (2017) 1--104.
\newblock \href {http://arxiv.org/abs/1705.11098} {\path{arXiv:1705.11098}}, \href {https://doi.org/10.1016/j.physrep.2017.06.001} {\path{doi:10.1016/j.physrep.2017.06.001}}.

\bibitem{Heisenberg:2018vsk}
L.~Heisenberg, {A systematic approach to generalisations of General Relativity and their cosmological implications}, Phys. Rept. 796 (2019) 1--113.
\newblock \href {http://arxiv.org/abs/1807.01725} {\path{arXiv:1807.01725}}, \href {https://doi.org/10.1016/j.physrep.2018.11.006} {\path{doi:10.1016/j.physrep.2018.11.006}}.

\bibitem{Clifton:2011jh}
T.~Clifton, P.~G. Ferreira, A.~Padilla, C.~Skordis, {Modified Gravity and Cosmology}, Phys. Rept. 513~(1-3) (2012) 1--189.
\newblock \href {http://arxiv.org/abs/1106.2476} {\path{arXiv:1106.2476}}, \href {https://doi.org/10.1016/j.physrep.2012.01.001} {\path{doi:10.1016/j.physrep.2012.01.001}}.

\bibitem{Olmo:2009xy}
G.~J. Olmo, H.~Sanchis-Alepuz, S.~Tripathi, {Dynamical aspects of generalized Palatini theories of gravity}, Phys. Rev. D 80 (2009) 024013.
\newblock \href {http://arxiv.org/abs/0907.2787} {\path{arXiv:0907.2787}}, \href {https://doi.org/10.1103/PhysRevD.80.024013} {\path{doi:10.1103/PhysRevD.80.024013}}.

\bibitem{Nojiri:2006ri}
S.~Nojiri, S.~D. Odintsov, {Introduction to modified gravity and gravitational alternative for dark energy}, Int. J. Geom. Meth. Mod. Phys. 4 (2007) 115--145.
\newblock \href {http://arxiv.org/abs/hep-th/0601213} {\path{arXiv:hep-th/0601213}}, \href {https://doi.org/10.1142/S0219887807001928} {\path{doi:10.1142/S0219887807001928}}.

\bibitem{Olmo:2011uz}
G.~J. Olmo, {Palatini Approach to Modified Gravity: f(R) Theories and Beyond}, Int. J. Mod. Phys. D 20 (2011) 413--462.
\newblock \href {http://arxiv.org/abs/1101.3864} {\path{arXiv:1101.3864}}, \href {https://doi.org/10.1142/S0218271811018925} {\path{doi:10.1142/S0218271811018925}}.

\bibitem{Zwiebach:1985uq}
B.~Zwiebach, {Curvature Squared Terms and String Theories}, Phys. Lett. B 156 (1985) 315--317.
\newblock \href {https://doi.org/10.1016/0370-2693(85)91616-8} {\path{doi:10.1016/0370-2693(85)91616-8}}.

\bibitem{Nojiri:2005jg}
S.~Nojiri, S.~D. Odintsov, {Modified Gauss--Bonnet theory as gravitational alternative for dark energy}, Phys. Lett. B 631 (2005) 1--6.
\newblock \href {http://arxiv.org/abs/hep-th/0508049} {\path{arXiv:hep-th/0508049}}, \href {https://doi.org/10.1016/j.physletb.2005.10.010} {\path{doi:10.1016/j.physletb.2005.10.010}}.

\bibitem{Banados:2010ix}
M.~Ba\~nados, P.~G. Ferreira, {Eddington's theory of gravity and its progeny}, Phys. Rev. Lett. 105 (2010) 011101.
\newblock \href {http://arxiv.org/abs/1006.1769} {\path{arXiv:1006.1769}}, \href {https://doi.org/10.1103/PhysRevLett.105.011101} {\path{doi:10.1103/PhysRevLett.105.011101}}.

\bibitem{Jimenez:2017uyn}
J.~B. Jim{\'e}nez, L.~Heisenberg, G.~J. Olmo, D.~Rubiera-Garcia, {Born--Infeld inspired modifications of gravity}, Phys. Rept. 727 (2018) 1--129.
\newblock \href {http://arxiv.org/abs/1704.03351} {\path{arXiv:1704.03351}}, \href {https://doi.org/10.1016/j.physrep.2017.11.001} {\path{doi:10.1016/j.physrep.2017.11.001}}.

\bibitem{Azri:2021det}
H.~Azri, K.~Y. Ekşi, C.~Karahan, S.~Nasri, {Ricci-determinant gravity: Dynamical aspects and astrophysical implications}, Phys. Rev. D 104~(6) (2021) 064049.
\newblock \href {http://arxiv.org/abs/2105.14378} {\path{arXiv:2105.14378}}, \href {https://doi.org/10.1103/PhysRevD.104.064049} {\path{doi:10.1103/PhysRevD.104.064049}}.

\bibitem{Bombaci:1997dd}
I.~Bombaci, Observational evidence for strange matter in compact objects from the x-ray burster 4u 1820-30, Phys. Rev. C 55 (1997) 1587--1590.
\newblock \href {https://doi.org/10.1103/PhysRevC.55.1587} {\path{doi:10.1103/PhysRevC.55.1587}}.

\bibitem{Glendenning:1995}
N.~K. Glendenning, {Prompt subsidence of a proto–neutron star into a black hole}, Astrophys. J. 448 (1995) 797--808.
\newblock \href {https://doi.org/10.1086/175999} {\path{doi:10.1086/175999}}.

\bibitem{Dai:1995}
Z.~G. Dai, Q.~H. Peng, T.~Lu, {The conversion of two-flavor to three-flavor quark matter in a supernova core}, Astrophys. J. 440 (1995) 815--822.
\newblock \href {https://doi.org/10.1086/175316} {\path{doi:10.1086/175316}}.

\bibitem{Cheng:1998wj}
K.~S. Cheng, Z.~G. Dai, T.~Lu, Strange stars and related astrophysical phenomena, Int. J. Mod. Phys. D 7 (1998) 139--172.
\newblock \href {http://arxiv.org/abs/astro-ph/9803034} {\path{arXiv:astro-ph/9803034}}, \href {https://doi.org/10.1142/S0218271898000127} {\path{doi:10.1142/S0218271898000127}}.

\bibitem{Chodos:1974je}
A.~Chodos, R.~L. Jaffe, K.~Johnson, C.~B. Thorn, {Baryon Structure in the Bag Theory}, Phys. Rev. D 10 (1974) 2599--2604.
\newblock \href {https://doi.org/10.1103/PhysRevD.10.2599} {\path{doi:10.1103/PhysRevD.10.2599}}.

\bibitem{Chodos:1974je1}
A.~Chodos, R.~L. Jaffe, K.~Johnson, C.~B. Thorn, V.~F. Weisskopf, {New extended model of hadrons}, Phys. Rev. D 9 (1974) 3471--3495.
\newblock \href {https://doi.org/10.1103/PhysRevD.9.3471} {\path{doi:10.1103/PhysRevD.9.3471}}.

\bibitem{Peshier:2000hx}
A.~Peshier, B.~K\"ampfer, G.~Soff, The equation of state of deconfined matter at finite chemical potential in a quasiparticle description, Phys. Rev. C 61 (2000) 045203.
\newblock \href {http://arxiv.org/abs/hep-ph/9911474} {\path{arXiv:hep-ph/9911474}}, \href {https://doi.org/10.1103/PhysRevC.61.045203} {\path{doi:10.1103/PhysRevC.61.045203}}.

\bibitem{Farhi:1984qu}
E.~Farhi, R.~L. Jaffe, {Strange Matter}, Phys. Rev. D 30 (1984) 2379--2390.
\newblock \href {https://doi.org/10.1103/PhysRevD.30.2379} {\path{doi:10.1103/PhysRevD.30.2379}}.

\bibitem{Alford:1998mk}
M.~G. Alford, K.~Rajagopal, F.~Wilczek, {Color-Flavor Locking and Chiral Symmetry Breaking in High Density QCD}, Nucl. Phys. B 537 (1999) 443--458.
\newblock \href {http://arxiv.org/abs/hep-ph/9804403} {\path{arXiv:hep-ph/9804403}}, \href {https://doi.org/10.1016/S0550-3213(98)00668-3} {\path{doi:10.1016/S0550-3213(98)00668-3}}.

\bibitem{Alford:1998mk1}
M.~G. Alford, K.~Rajagopal, F.~Wilczek, {Color-Flavor Locking and Chiral Symmetry Breaking in High Density QCD}, Nucl. Phys. B 537 (1999) 443--458.
\newblock \href {http://arxiv.org/abs/hep-ph/9804403} {\path{arXiv:hep-ph/9804403}}, \href {https://doi.org/10.1016/S0550-3213(98)00668-3} {\path{doi:10.1016/S0550-3213(98)00668-3}}.

\bibitem{Flores:2017nonradial}
C.~V. Flores, Z.~B. Hall, P.~Jaikumar, {Nonradial oscillation modes of compact stars with a crust}, Phys. Rev. C 96~(6) (2017) 065803.
\newblock \href {http://arxiv.org/abs/1708.05991} {\path{arXiv:1708.05991}}, \href {https://doi.org/10.1103/PhysRevC.96.065803} {\path{doi:10.1103/PhysRevC.96.065803}}.

\bibitem{Alford:2005js}
M.~Alford, M.~Braby, M.~Paris, S.~Reddy, Hybrid stars that masquerade as neutron stars, Astrophys. J. 629~(2) (2005) 969--978.
\newblock \href {http://arxiv.org/abs/nucl-th/0411016} {\path{arXiv:nucl-th/0411016}}, \href {https://doi.org/10.1086/430902} {\path{doi:10.1086/430902}}.

\bibitem{Asbell:2017}
J.~Asbell, P.~Jaikumar, Oscillation modes of strange quark stars with a strangelet crust, J. Phys. Conf. Ser. 861 (2017) 012029.
\newblock \href {https://doi.org/10.1088/1742-6596/861/1/012029} {\path{doi:10.1088/1742-6596/861/1/012029}}.

\bibitem{Zhang:2021unified}
C.~Zhang, R.~B. Mann, Unified interacting quark matter and its astrophysical implications, Phys. Rev. D 103~(6) (2021) 063018.
\newblock \href {http://arxiv.org/abs/2101.00000} {\path{arXiv:2101.00000}}, \href {https://doi.org/10.1103/PhysRevD.103.063018} {\path{doi:10.1103/PhysRevD.103.063018}}.

\bibitem{Zhang:2021stellar}
C.~Zhang, M.~Gammon, R.~B. Mann, Stellar structure and stability of charged interacting quark stars and their scaling behaviour (2021).
\newblock \href {http://arxiv.org/abs/2108.13972} {\path{arXiv:2108.13972}}.

\bibitem{Holdom:2018hfd}
B.~Holdom, J.~Ren, C.~Zhang, {Quark Matter May Not Be Strange}, Phys. Rev. Lett. 120~(22) (2018) 222001.
\newblock \href {http://arxiv.org/abs/1804.XXXX} {\path{arXiv:1804.XXXX}}, \href {https://doi.org/10.1103/PhysRevLett.120.222001} {\path{doi:10.1103/PhysRevLett.120.222001}}.

\bibitem{Miao:2021xuq}
Z.~Miao, J.-L. Jiang, A.~Li, L.-W. Chen, {Bayesian inference of strange star equation of state using the GW170817 and GW190425 data}, Astrophys. J. Lett. 917~(2) (2021) L22.
\newblock \href {http://arxiv.org/abs/2105.11768} {\path{arXiv:2105.11768}}, \href {https://doi.org/10.3847/2041-8213/ac0ffb} {\path{doi:10.3847/2041-8213/ac0ffb}}.

\bibitem{Oikonomou:2023cflqs}
P.~T. Oikonomou, C.~C. Moustakidis, Color-flavor locked quark stars in light of the compact object in the hess j1731-347 and the gw190814 event, Phys. Rev. D 108~(6) (2023) 063010.
\newblock \href {http://arxiv.org/abs/2307.03918} {\path{arXiv:2307.03918}}, \href {https://doi.org/10.1103/PhysRevD.108.063010} {\path{doi:10.1103/PhysRevD.108.063010}}.

\bibitem{Lopes:2022nature}
L.~L. Lopes, D.~P. Menezes, {On the Nature of the Mass-Gap Object in the GW190814 Event}, Astrophys. J. 936~(1) (2022) 41.
\newblock \href {http://arxiv.org/abs/2207.XXXXX} {\path{arXiv:2207.XXXXX}}, \href {https://doi.org/10.3847/1538-4357/ac81c1} {\path{doi:10.3847/1538-4357/ac81c1}}.

\bibitem{BecerraVergara:2019wgj}
E.~A. Becerra-Vergara, S.~Mojica, F.~D. Lora-Clavijo, A.~Cruz-Osorio, Anisotropic quark stars with an interacting quark equation of state, Phys. Rev. D 100~(10) (2019) 103006.
\newblock \href {http://arxiv.org/abs/1907.11838} {\path{arXiv:1907.11838}}, \href {https://doi.org/10.1103/PhysRevD.100.103006} {\path{doi:10.1103/PhysRevD.100.103006}}.

\bibitem{Gammon:2024quark}
M.~Gammon, S.~Rourke, R.~B. Mann, {Quark stars with a unified interacting equation of state in regularized 4D Einstein-Gauss-Bonnet gravity}, Phys. Rev. D 109~(2) (2024) 024026.
\newblock \href {https://doi.org/10.1103/PhysRevD.109.024026} {\path{doi:10.1103/PhysRevD.109.024026}}.

\bibitem{Banerjee:2021strange}
A.~Banerjee, T.~Tangphati, P.~Channuie, {Strange quark stars in 4D Einstein--Gauss--Bonnet gravity}, Astrophys. J. 909~(1) (2021) 14.
\newblock \href {http://arxiv.org/abs/2102.12508} {\path{arXiv:2102.12508}}, \href {https://doi.org/10.3847/1538-4357/abd9f2} {\path{doi:10.3847/1538-4357/abd9f2}}.

\bibitem{Banerjee:2020xxx}
A.~Banerjee, T.~Tangphati, D.~Samart, P.~Channuie, {Quark stars in 4D Einstein--Gauss--Bonnet gravity with an interacting quark equation of state}, Astrophys. J. 906~(2) (2021) 114.
\newblock \href {http://arxiv.org/abs/2007.06547} {\path{arXiv:2007.06547}}, \href {https://doi.org/10.3847/1538-4357/abc8e2} {\path{doi:10.3847/1538-4357/abc8e2}}.

\bibitem{Pretel:2022electrically}
J.~M.~Z. Pretel, A.~Banerjee, A.~Pradhan, {Electrically charged quark stars in 4D Einstein--Gauss--Bonnet gravity}, Eur. Phys. J. C 82~(2) (2022) 180.
\newblock \href {http://arxiv.org/abs/2201.XXXXX} {\path{arXiv:2201.XXXXX}}, \href {https://doi.org/10.1140/epjc/s10052-022-10166-7} {\path{doi:10.1140/epjc/s10052-022-10166-7}}.

\bibitem{Rincon:2023anisotropic}
{\'A}.~Rinc{\'o}n, G.~Panotopoulos, I.~Lopes, {Anisotropic quark stars with an interacting quark equation of state within the complexity factor formalism}, Universe 9~(2) (2023) 72.
\newblock \href {http://arxiv.org/abs/2301.XXXXX} {\path{arXiv:2301.XXXXX}}, \href {https://doi.org/10.3390/universe9020072} {\path{doi:10.3390/universe9020072}}.

\bibitem{Kim:2021neutronHL}
K.~Kim, J.~J. Oh, C.~Park, E.~J. Son, Neutron star structure in ho{\v{r}}ava-lifshitz gravity, Phys. Rev. D 103~(4) (2021) 044052.
\newblock \href {http://arxiv.org/abs/2101.XXXX} {\path{arXiv:2101.XXXX}}, \href {https://doi.org/10.1103/PhysRevD.103.044052} {\path{doi:10.1103/PhysRevD.103.044052}}.

\bibitem{Das:2025xxx}
K.~P. Das, U.~Debnath, {Study of stable dark energy stars in Ho\v{r}ava--Lifshitz gravity}, Eur. Phys. J. C 85~(3) (2025) 329.
\newblock \href {https://doi.org/10.1140/epjc/s10052-025-XXXXX-X} {\path{doi:10.1140/epjc/s10052-025-XXXXX-X}}.

\bibitem{Azri:2023detT}
H.~Azri, S.~Nasri, {Gravity from the determinant of the energy-momentum: Astrophysical implications}, Phys. Lett. B 836 (2023) 137626.
\newblock \href {http://arxiv.org/abs/2211.XXXX} {\path{arXiv:2211.XXXX}}, \href {https://doi.org/10.1016/j.physletb.2022.137626} {\path{doi:10.1016/j.physletb.2022.137626}}.

\bibitem{Das:2025possible}
K.~P. Das, U.~Debnath, {Possible formation of Chaplygin dark star in gravity from the determinant of the energy--momentum tensor}, Phys. Dark Univ. 48 (2025) 101959.
\newblock \href {https://doi.org/10.1016/j.dark.2025.101959} {\path{doi:10.1016/j.dark.2025.101959}}.

\bibitem{Mota:2024xxx}
C.~E. Mota, J.~M.~Z. Pretel, C.~O.~V. Flores, {Neutron stars in $f(R,L_m,T)$ gravity}, Eur. Phys. J. C 84~(7) (2024) 673.
\newblock \href {http://arxiv.org/abs/2404.XXXXX} {\path{arXiv:2404.XXXXX}}, \href {https://doi.org/10.1140/epjc/s10052-024-13036-5} {\path{doi:10.1140/epjc/s10052-024-13036-5}}.

\bibitem{Miller:2021c}
M.~C. Miller, F.~K. Lamb, A.~J. Dittmann, et~al., {The radius of PSR~J0740+6620 from NICER and XMM-Newton data}, Astrophys. J. Lett. 918 (2021) L28.
\newblock \href {http://arxiv.org/abs/2105.06979} {\path{arXiv:2105.06979}}, \href {https://doi.org/10.3847/2041-8213/ac089b} {\path{doi:10.3847/2041-8213/ac089b}}.

\bibitem{Abbott:2017vtc}
B.~P. Abbott, et~al., {GW170817: Observation of Gravitational Waves from a Binary Neutron Star Inspiral}, Phys. Rev. Lett. 119~(16) (2017) 161101.
\newblock \href {http://arxiv.org/abs/1710.05832} {\path{arXiv:1710.05832}}, \href {https://doi.org/10.1103/PhysRevLett.119.161101} {\path{doi:10.1103/PhysRevLett.119.161101}}.

\bibitem{Bodmer:1971we}
A.~R. Bodmer, {Collapsed Nuclei}, Phys. Rev. D 4 (1971) 1601--1606.
\newblock \href {https://doi.org/10.1103/PhysRevD.4.1601} {\path{doi:10.1103/PhysRevD.4.1601}}.

\bibitem{Witten:1984rs}
E.~Witten, {Cosmic Separation of Phases}, Phys. Rev. D 30 (1984) 272--285.
\newblock \href {https://doi.org/10.1103/PhysRevD.30.272} {\path{doi:10.1103/PhysRevD.30.272}}.

\bibitem{Terazawa:1979}
H.~Terazawa, {Quark Matter and Quark Stars}, in: {2nd KEK Symposium on Radiation Dosimetry}, Tsukuba, Japan, 1979, pp. 22--23.

\bibitem{Holdom:2017iqi}
B.~Holdom, J.~Ren, C.~Zhang, {Quark Matter May Not Be Strange}, Phys. Rev. Lett. 120~(22) (2018) 222001.
\newblock \href {http://arxiv.org/abs/1707.06610} {\path{arXiv:1707.06610}}, \href {https://doi.org/10.1103/PhysRevLett.120.222001} {\path{doi:10.1103/PhysRevLett.120.222001}}.

\bibitem{Fraga:2001id}
E.~S. Fraga, R.~D. Pisarski, J.~Schaffner-Bielich, {Small, dense quark stars from perturbative QCD}, Phys. Rev. D 63 (2001) 121702.
\newblock \href {http://arxiv.org/abs/hep-ph/0101143} {\path{arXiv:hep-ph/0101143}}, \href {https://doi.org/10.1103/PhysRevD.63.121702} {\path{doi:10.1103/PhysRevD.63.121702}}.

\bibitem{Fraga:2013qra}
E.~S. Fraga, A.~Kurkela, A.~Vuorinen, {Interacting quark matter equation of state for compact stars}, Astrophys. J. Lett. 781 (2014) L25.
\newblock \href {http://arxiv.org/abs/1311.5154} {\path{arXiv:1311.5154}}, \href {https://doi.org/10.1088/2041-8205/781/2/L25} {\path{doi:10.1088/2041-8205/781/2/L25}}.

\bibitem{Rajagopal:2000ff}
K.~Rajagopal, F.~Wilczek, {Enforced electrical neutrality of the color-flavor locked phase}, Phys. Rev. Lett. 86 (2001) 3492--3495.
\newblock \href {http://arxiv.org/abs/hep-ph/0012039} {\path{arXiv:hep-ph/0012039}}, \href {https://doi.org/10.1103/PhysRevLett.86.3492} {\path{doi:10.1103/PhysRevLett.86.3492}}.

\bibitem{Lugones:2002iz}
G.~Lugones, J.~E. Horvath, {Color flavor locked strange matter}, Phys. Rev. D 66 (2002) 074017.
\newblock \href {http://arxiv.org/abs/hep-ph/0211070} {\path{arXiv:hep-ph/0211070}}, \href {https://doi.org/10.1103/PhysRevD.66.074017} {\path{doi:10.1103/PhysRevD.66.074017}}.

\bibitem{Zhang:2020qnw}
C.~Zhang, R.~B. Mann, {Unified Interacting Quark Matter and its Astrophysical Implications}, Phys. Rev. D 103~(6) (2021) 063018.
\newblock \href {http://arxiv.org/abs/2009.07182} {\path{arXiv:2009.07182}}, \href {https://doi.org/10.1103/PhysRevD.103.063018} {\path{doi:10.1103/PhysRevD.103.063018}}.

\bibitem{Zhang:2021xxx}
C.~Zhang, {Gravitational wave echoes from interacting quark stars}, Phys. Rev. D 104~(8) (2021) 083032.
\newblock \href {http://arxiv.org/abs/2107.09654} {\path{arXiv:2107.09654}}, \href {https://doi.org/10.1103/PhysRevD.104.083032} {\path{doi:10.1103/PhysRevD.104.083032}}.

\bibitem{Blaschke:2022crossover}
D.~Blaschke, E.-O. Hanu, S.~Liebing, {Neutron stars with crossover to color superconducting quark matter}, Phys. Rev. C 105 (2022) 035804.
\newblock \href {http://arxiv.org/abs/2112.12145} {\path{arXiv:2112.12145}}, \href {https://doi.org/10.1103/PhysRevC.105.035804} {\path{doi:10.1103/PhysRevC.105.035804}}.

\bibitem{Blaschke:2022}
D.~Blaschke, U.~Shukla, O.~Ivanytskyi, S.~Liebing, {Effect of color superconductivity on the mass of hybrid neutron stars in an effective model with perturbative QCD asymptotics}, Phys. Rev. D 107~(6) (2023) 063034.
\newblock \href {http://arxiv.org/abs/2212.14856} {\path{arXiv:2212.14856}}, \href {https://doi.org/10.1103/PhysRevD.107.063034} {\path{doi:10.1103/PhysRevD.107.063034}}.

\bibitem{Gammon:2023xxx}
M.~Gammon, S.~Rourke, R.~B. Mann, {Quark stars with a unified interacting equation of state in regularized 4D Einstein-Gauss-Bonnet gravity}, Phys. Rev. D 109~(2) (2024) 024026.
\newblock \href {http://arxiv.org/abs/2309.00703} {\path{arXiv:2309.00703}}, \href {https://doi.org/10.1103/PhysRevD.109.024026} {\path{doi:10.1103/PhysRevD.109.024026}}.

\bibitem{Tangphati:2023}
T.~Tangphati, D.~J. Gogoi, A.~Pradhan, A.~Banerjee, {Investigating stable quark stars in Rastall-Rainbow gravity and their compatibility with gravitational wave observations}, JHEAp 42 (2024) 12--26.
\newblock \href {http://arxiv.org/abs/2311.16869} {\path{arXiv:2311.16869}}, \href {https://doi.org/10.1016/j.jheap.2024.02.001} {\path{doi:10.1016/j.jheap.2024.02.001}}.

\bibitem{Yuan:2023tweflavor}
W.-L. Yuan, A.~Li, {Two-flavor Color Superconducting Quark Stars May Not Exist}, Astrophys. J. 966~(1) (2024) 3.
\newblock \href {http://arxiv.org/abs/2312.17102} {\path{arXiv:2312.17102}}, \href {https://doi.org/10.3847/1538-4357/ad2f4e} {\path{doi:10.3847/1538-4357/ad2f4e}}.

\bibitem{Das:2024iqs}
K.~P. Das, P.~Bhar, U.~Debnath, {Anisotropic quark stars with an interacting quark equation of state in extra dimension}, Eur. Phys. J. C 84 (2024) 952.
\newblock \href {https://doi.org/10.1140/epjc/s10052-024-13234-2} {\path{doi:10.1140/epjc/s10052-024-13234-2}}.

\bibitem{Das:2024cflstrangestars}
K.~P. Das, A.~Karmakar, U.~Debnath, {Color flavor locked strange stars in de Rham–Gabadadze–Tolley like massive gravity}, Eur. Phys. J. C 84 (2024) 1213.
\newblock \href {https://doi.org/10.1140/epjc/s10052-024-13556-1} {\path{doi:10.1140/epjc/s10052-024-13556-1}}.

\bibitem{Komathiraj:2007fw}
K.~Komathiraj, S.~D. Maharaj, {Analytical models for quark stars}, Int. J. Mod. Phys. D 16 (2007) 1803--1811.
\newblock \href {http://arxiv.org/abs/0712.1278} {\path{arXiv:0712.1278}}, \href {https://doi.org/10.1142/S0218271807010825} {\path{doi:10.1142/S0218271807010825}}.

\bibitem{Harko:2002}
T.~Harko, M.~K. Mak, {An Exact Anisotropic Quark Star Model}, Chin. J. Astron. Astrophys. 2 (2002) 248--259.
\newblock \href {http://arxiv.org/abs/gr-qc/0204015} {\path{arXiv:gr-qc/0204015}}, \href {https://doi.org/10.1088/1009-9271/2/3/08} {\path{doi:10.1088/1009-9271/2/3/08}}.

\bibitem{Deb:2016nuv}
D.~Deb, S.~V. Ketov, P.~H. R.~S. Moraes, S.~K. Maurya, S.~Ray, {Relativistic model for anisotropic strange stars}, Annals Phys. 387 (2017) 239--252.
\newblock \href {http://arxiv.org/abs/1606.00713} {\path{arXiv:1606.00713}}, \href {https://doi.org/10.1016/j.aop.2017.10.011} {\path{doi:10.1016/j.aop.2017.10.011}}.

\bibitem{Das:2023iqs}
H.~C. Das, L.~L. Lopes, {Anisotropic strange stars in the spotlight: unveiling constraints through observational data}, Mon. Not. Roy. Astron. Soc. 525 (2023) 3571--3585.
\newblock \href {http://arxiv.org/abs/2306.00326} {\path{arXiv:2306.00326}}, \href {https://doi.org/10.1093/mnras/stad2587} {\path{doi:10.1093/mnras/stad2587}}.

\bibitem{Lopes:2023xxx}
L.~L. Lopes, H.~C. Das, {Spherically symmetric anisotropic strange stars}, Eur. Phys. J. C 84~(2) (2024) 166.
\newblock \href {http://arxiv.org/abs/2312.00310} {\path{arXiv:2312.00310}}, \href {https://doi.org/10.1140/epjc/s10052-024-12532-7} {\path{doi:10.1140/epjc/s10052-024-12532-7}}.

\bibitem{Pretel:2023xxx}
J.~M.~Z. Pretel, T.~Tangphati, A.~Banerjee, A.~Pradhan, {Effects of anisotropic pressure on interacting quark star structure}, Phys. Lett. B 848 (2024) 138375.
\newblock \href {http://arxiv.org/abs/2311.18770} {\path{arXiv:2311.18770}}, \href {https://doi.org/10.1016/j.physletb.2023.138375} {\path{doi:10.1016/j.physletb.2023.138375}}.

\bibitem{Asbell:2017vdd}
J.~Asbell, P.~Jaikumar, {Oscillation modes of strange quark stars with a strangelet crust}, J. Phys. Conf. Ser. 861 (2017) 012029.
\newblock \href {https://doi.org/10.1088/1742-6596/861/1/012029} {\path{doi:10.1088/1742-6596/861/1/012029}}.

\bibitem{BecerraVergara:2019yjj}
E.~Becerra-Vergara, S.~Mojica, F.~D. Lora-Clavijo, A.~Cruz-Osorio, {Anisotropic quark stars with an interacting quark equation of state}, Phys. Rev. D 100~(10) (2019) 103006.
\newblock \href {http://arxiv.org/abs/1909.02004} {\path{arXiv:1909.02004}}, \href {https://doi.org/10.1103/PhysRevD.100.103006} {\path{doi:10.1103/PhysRevD.100.103006}}.

\bibitem{Beringer:2012}
J.~Beringer, J.-F. Arguin, R.~M. Barnett, K.~Copic, O.~Dahl, D.~E. Groom, C.-J. Lin, J.~Lys, H.~Murayama, C.~G. Wohl, et~al., {Review of Particle Physics}, Phys. Rev. D 86~(1) (2012) 010001.
\newblock \href {https://doi.org/10.1103/PhysRevD.86.010001} {\path{doi:10.1103/PhysRevD.86.010001}}.

\bibitem{Burgio:2018}
G.~F. Burgio, A.~F. Fantina, {Nuclear Equation of State for Compact Stars and Supernovae}, in: The Physics and Astrophysics of Neutron Stars, Vol. 457 of Astrophysics and Space Science Library, Springer, Cham, 2018, pp. 255--279.
\newblock \href {https://doi.org/10.1007/978-3-319-97616-7_8} {\path{doi:10.1007/978-3-319-97616-7_8}}.

\bibitem{Blaschke:2018}
D.~Blaschke, N.~Chamel, {Phases of Dense Matter in Compact Stars}, Astrophys. Space Sci. Libr. 457 (2018) 337.
\newblock \href {https://doi.org/10.1007/978-3-319-97616-7_8} {\path{doi:10.1007/978-3-319-97616-7_8}}.

\bibitem{Fraga:2000ew}
E.~S. Fraga, R.~D. Pisarski, J.~Schaffner-Bielich, {Small, dense quark stars from perturbative QCD}, Phys. Rev. D 63~(12) (2001) 121702.
\newblock \href {http://arxiv.org/abs/hep-ph/0101143} {\path{arXiv:hep-ph/0101143}}, \href {https://doi.org/10.1103/PhysRevD.63.121702} {\path{doi:10.1103/PhysRevD.63.121702}}.

\bibitem{Chandrasekhar:1964zz}
S.~Chandrasekhar, {Dynamical Instability of Gaseous Masses Approaching the Schwarzschild Limit in General Relativity}, Phys. Rev. Lett. 12 (1964) 114--116.
\newblock \href {https://doi.org/10.1103/PhysRevLett.12.114} {\path{doi:10.1103/PhysRevLett.12.114}}.

\bibitem{Chan:1993}
R.~Chan, L.~Herrera, N.~O. Santos, {Dynamical instability for radiating anisotropic collapse}, Mon. Not. Roy. Astron. Soc. 265 (1993) 533--544.
\newblock \href {https://doi.org/10.1093/mnras/265.3.533} {\path{doi:10.1093/mnras/265.3.533}}.

\bibitem{Glass:1983stability}
E.~N. Glass, A.~Harpaz, {The stability of relativistic gas spheres}, Mon. Not. Roy. Astron. Soc. 202 (1983) 159--171.
\newblock \href {https://doi.org/10.1093/mnras/202.1.159} {\path{doi:10.1093/mnras/202.1.159}}.

\bibitem{Moustakidis:2017tlk}
C.~C. Moustakidis, {The stability of relativistic stars and the role of the adiabatic index}, Gen. Rel. Grav. 49~(5) (2017) 68.
\newblock \href {http://arxiv.org/abs/1612.01726} {\path{arXiv:1612.01726}}, \href {https://doi.org/10.1007/s10714-017-2220-3} {\path{doi:10.1007/s10714-017-2220-3}}.

\bibitem{Buchdahl:1959zz}
H.~A. Buchdahl, {General Relativistic Fluid Spheres}, Phys. Rev. 116 (1959) 1027--1034.
\newblock \href {https://doi.org/10.1103/PhysRev.116.1027} {\path{doi:10.1103/PhysRev.116.1027}}.

\bibitem{Jiang:2019xwn}
R.~Jiang, D.~Wen, H.~Chen, {Universal behavior of a compact star based upon the gravitational binding energy}, Phys. Rev. D 100~(12) (2019) 123010.
\newblock \href {http://arxiv.org/abs/1911.10935} {\path{arXiv:1911.10935}}, \href {https://doi.org/10.1103/PhysRevD.100.123010} {\path{doi:10.1103/PhysRevD.100.123010}}.

\bibitem{Bagchi:2011}
M.~Bagchi, {The role of binding energy of neutron stars on the accretion driven evolution}, Mon. Not. Roy. Astron. Soc. Lett. 413 (2011) L47--L50.
\newblock \href {http://arxiv.org/abs/1101.2006} {\path{arXiv:1101.2006}}, \href {https://doi.org/10.1111/j.1745-3933.2011.01038.x} {\path{doi:10.1111/j.1745-3933.2011.01038.x}}.

\end{thebibliography}

\end{document}